\documentclass[aps,prb,reprint,superscriptaddress]{revtex4-2}
\usepackage{graphicx} 
\usepackage{amsmath}
\usepackage{siunitx}
\usepackage{pifont}
\usepackage{booktabs}

\newcommand{\cmark}{\ding{51}}
\newcommand{\xmark}{\ding{55}}

\begin{document}

\title{Ab initio study of magnetic Cu$_2$Sb compounds related to antiferromagnetic CuMnAs} 

\author{Vojt\v{e}ch Pa\v{r}\'izek}
\affiliation{Institute of Physics, Czech Academy of Sciences, Cukrovarnick\'{a} 10, 162 00 Praha 6 Czech Republic}
\author{Jakub \v{Z}elezn\'{y}}
\affiliation{Institute of Physics, Czech Academy of Sciences, Cukrovarnick\'{a} 10, 162 00 Praha 6 Czech Republic}

\begin{abstract}
The antiferromagnetic metal CuMnAs has become a workhorse of antiferromagnetic spintronics. A broad range of phenomena has been demonstrated in this material, most notably current-induced switching and heat-driven quench switching, both first realized in CuMnAs. CuMnAs belongs to a much larger family of
intermetallic compounds sharing the tetragonal Cu$_2$Sb-type structure, yet the
magnetic properties of most of its members, and their potential for spintronics,
have not been studied. Here we report a systematic
density functional theory (DFT) study of the magnetic and electronic structure
of more than 50 compounds of this family, intended to guide future theoretical
and experimental work toward spintronic applications. We find a wide range of
magnetic ground states, among them over 20 new antiferromagnetic candidates.

\end{abstract}

\maketitle

\section{Introduction}

Antiferromagnets are magnetically ordered materials with no or negligible net magnetic moment. Although they are common in nature and have been known for a long time, they have only gained significant attention from the research community in recent years; a development closely tied to the field of spintronics, which aims to exploit the electron spin in microelectronics. Their lack of a net moment makes them unsuitable for the conventional applications of ferromagnets and makes them difficult to study, since the magnetic order is hard to detect and hard to manipulate with magnetic fields. In spintronics, however, the key ingredient is the interaction between the electron spin and the magnetic order, which does not rely on a net moment. Over the last decade, many spintronic phenomena and functionalities have been realized in antiferromagnets, clearly establishing their potential. This progress is driven both by fundamental interest and by the potential advantages of antiferromagnets, such as their very fast magnetic dynamics compared to ferromagnets.
 
CuMnAs, a collinear antiferromagnet with the tetragonal Cu$_2$Sb crystal structure, has been a key material for the development of antiferromagnetic spintronics~\cite{wadley2013}. It was the first antiferromagnet in which electrical switching was demonstrated~\cite{wadley2016}, utilizing the so-called spin-orbit torque, a current-induced torque arising from spin-orbit coupling and broken inversion symmetry. In CuMnAs the inversion symmetry is broken locally at each magnetic sublattice, leading to a staggered torque that efficiently couples to the magnetic order~\cite{Zelezny2017}. Following this pivotal experiment, CuMnAs has become a workhorse material for the field, with a broad range of spintronic phenomena demonstrated: current-induced switching of domain walls~\cite{grzybowski2017}, multi-level memory capability~\cite{olejnik2017}, terahertz switching~\cite{olejnik2018}, optical readout~\cite{saidl2017,schmid2026}, ultrasharp atomic domain walls~\cite{krizek2022}, domain engineering through strain~\cite{reimers2024}, gating~\cite{grzybowski2019}, electrically induced and detected reversal of the N\'eel vector~\cite{godinho2018}, nanoscale scanning N\'eel vector reading using photocurrent~\cite{schmid2026}, and, remarkably, switching driven not by current but by heat~\cite{kaspar2020,zubac2021}.
 
This last effect, known as quench switching, is particularly notable. The heat can be delivered either by electrical current or by optical pulses, and the switching is associated with a breakdown of the domain structure, although its microscopic origin remains poorly understood. Crucially, it can produce large and highly reproducible resistance changes—around 20\% at room temperature and 100\% at low temperatures—with regular, multi-level behavior whose relaxation follows the Kohlrausch stretched exponential. These features may make it attractive for neuromorphic computing~\cite{surynek2025,zubac2025}. Significantly, quench switching has recently been demonstrated in another member of the Cu$_2$Sb family, Mn$_2$As~\cite{olejnik2025}, where the resistance change reaches 700\% at low temperatures. Mn$_2$As is also a collinear antiferromagnet but possesses a magnetic structure distinct from that of CuMnAs. That the same effect appears in a structurally related yet magnetically different compound suggests it may be a more general feature of this class of materials.
 
Many other magnetic materials in the Cu$_2$Sb family are known, exhibiting a variety of magnetic orders. CuMnAs, Mn$_2$As~\cite{austin1962}, Cr$_2$As~\cite{yamaguchi1972}, Fe$_2$As~\cite{katsuraki1966}, and CuFeAs~\cite{thakur2014} are collinear antiferromagnets with a range of different orderings, whereas Mn$_2$Sb is ferrimagnetic~\cite{wilkinson1957} and CuFeSb is ferromagnetic~\cite{qian2012}. Alloys such as (Cr$_{1-x}$Mn$_x$)$_2$As have also been investigated~\cite{yamaguchi1999}. Many of these compounds have high Curie or N\'eel temperatures, making them potentially attractive for applications. Still other materials containing magnetic elements have been synthesized, but their magnetic order has not been studied experimentally.
 
The Cu$_2$Sb intermetallic compounds thus form a large family of materials exhibiting a variety of magnetic orders, with potential for spintronics applications. By combining different elements or by alloying, it may be possible to engineer desirable properties. However, most of these materials have been studied only in a limited way, or not at all. Apart from CuMnAs, only a few, including Cr$_2$As~\cite{shirai1993}, Mn$_2$As, Fe$_2$As~\cite{Zhang2013}, and Mn$_2$Sb~\cite{suzuki1992}, have been studied theoretically.
 
To bridge this gap, we use \textit{ab initio} calculations to study the electronic and magnetic structure of a large number of materials from this class. We consider combinations of Mn, Fe, Cr, Co, Ga, Cu, and Ni with As or Sb, including both experimentally synthesized compounds and hypothetical materials that have not yet been realized. The stability of the hypothetical materials must ultimately be determined experimentally, but we expect many of them to be realizable, given the close chemical similarity within this family. We note that epitaxial growth, for example by molecular beam epitaxy, can stabilize the tetragonal phase: CuMnAs, for instance, is orthorhombic in bulk but tetragonal when grown as thin films by molecular beam epitaxy. For each material we determine the magnetic structure by computing the energies of the magnetic configurations most common in this family, and we then calculate the corresponding electronic structure. Our work is intended to serve as a basis for further theoretical and experimental studies of this family of materials.

\section{Workflow} \label{sec:workflow}

DFT calculations were performed using the OpenMX code. Workflow automation and data management were facilitated by AiiDA, a framework that enables automated execution of calculations and storage of calculation inputs and outputs in a database. To integrate OpenMX with AiiDA, we developed a dedicated plugin, available at~\cite{aiida_openmx_plugin_2026}.

\begin{figure}[ht]
\centering
\includegraphics[width=0.483\textwidth,page=1]{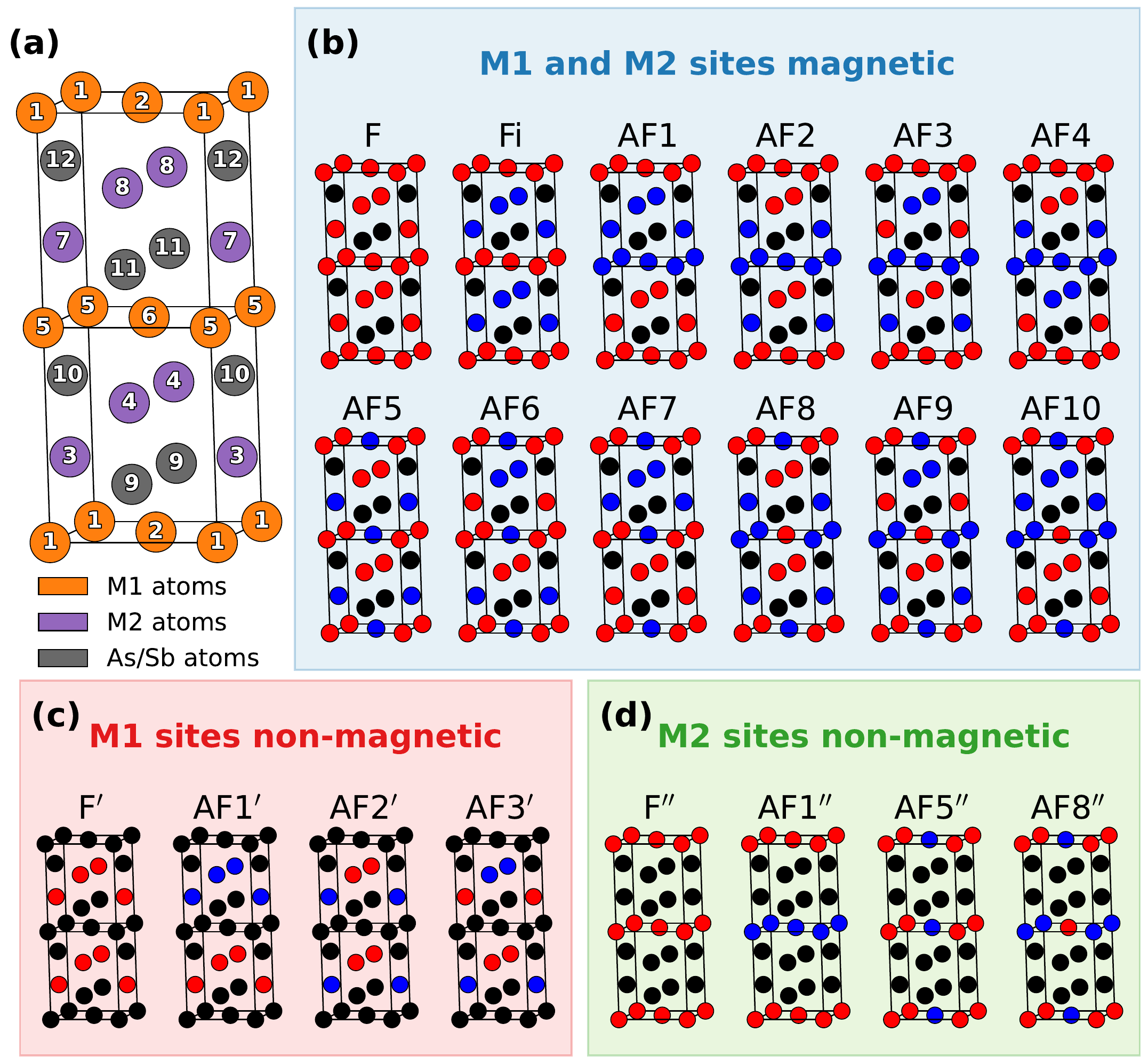}
\caption{(a) The Cu$_2$Sb structure doubled along the $c$ axis, denoting the M1 and M2 sites and atom indices as used within this work. (b) Magnetic structures of Cu$_2$Sb-type materials considered in this work as proposed by~\cite{Zhang2013}. In the case where the M1 site is non-magnetic or the M2 site is non-magnetic, this reduces to 4 possible structures, shown in (c) and (d), respectively.}
\label{fig:magnetic structures}
\end{figure}

To determine the magnetic structure we perform calculations of different magnetic orders and compare their total energies. Cu$_2$Sb-type structures are known to favor collinear arrangements of magnetic moments, either within the primitive cell or with a doubling along the $z$ direction~\cite{Fruchart2005}. There are two magnetic sites in the Cu$_2$Sb structure, which we denote by M1 and M2 as shown in Fig.~\ref{fig:magnetic structures}(a). By considering the nearest- and next-nearest-neighbor interactions between M1--M1, M1--M2, and M2--M2 atoms, and assuming collinear magnetic order with possible doubling along the $z$ direction, Zhang \textit{et al.} identified twelve possible magnetic structures~\cite{Zhang2013}: one ferromagnetic configuration (F1), one ferrimagnetic configuration (Fi1), and ten antiferromagnetic configurations (AFi), as shown in Fig.~\ref{fig:magnetic structures}(b). We consider this set of magnetic structures in the present work. In the case where only one of the sites is magnetic, this reduces to four magnetic structures (see Fig.~\ref{fig:magnetic structures}(c),(d)). Note that although we always consider these sets of magnetic structures, some of them do not converge and are therefore not included in the results. Generally, we expect that the issues with convergence are because the magnetic structure is not realistic for the material and thus the unconverged calculations should not influence the ground state determination.

Since the lattice parameters for most of the materials we have studied are not known, we first optimized the lattice constants and the internal positions while preserving the P4/nmm space group. This amounts to optimizing the $a$ and $c$ lattice constants and the $z$ coordinates of the M2 and As/Sb sites. We note that for consistency we optimized the lattice even for materials where the lattice parameters are experimentally known. We utilized the variable-cell optimization (VCO) for lattice optimization. The VCO calculations were performed using the AF5 magnetic structure, as the explicit inclusion of magnetic order during the optimization was found to improve the agreement of the resulting lattice parameters and atomic positions with experimental data. Several VCO methods available in OpenMX were tested, as described in Sec.~\ref{sec:methods}. Additionally, in materials that contain two different transition metal elements, it is not clear which one will occupy the M1 site and which one the M2 site. Thus we calculate both variants. Although only the lower energy is expected to be realized, we include both sets of calculations for completeness. 

The optimized unit cell was subsequently doubled along the $z$ axis, and the magnetic structures shown in Fig.~\ref{fig:magnetic structures} were used as input configurations. For materials containing Cu or Ga on the M1 or M2 site, the structures from the red or green box in Fig.~\ref{fig:magnetic structures}, respectively, were considered. For materials containing neither Cu nor Ga on these sites, the structures from the blue box were used. Consequently, either 4 or 12 calculations were performed for each material. This stage consisted of two consecutive DFT calculations: an initial calculation with an electronic temperature of \qty{2000}{\kelvin}, followed by a calculation at \qty{300}{\kelvin}. The elevated electronic temperature was employed to achieve convergence in a smaller number of iterations. In addition to the standard DFT calculations, a series of DFT+$U$ calculations was performed using element-dependent values of the on-site Coulomb interaction parameter $U$. The correction was applied only to the transition-metal species: $U=\qty{3}{\electronvolt}$ for Fe, $U=\qty{3.5}{\electronvolt}$ for Co, Cr, and Mn, and $U=\qty{4}{\electronvolt}$ for Ni. For As, Sb, Cu, and Ga, no Hubbard correction was applied, corresponding to $U=\qty{0}{\electronvolt}$. After convergence, the calculation with the lowest total energy for each material was selected for the density of states (DOS) and band structure calculations presented in the Supplementary Material.

\section{Results}

\begin{figure}[ht]
\centering
\includegraphics[width=0.40\textwidth,page=1]{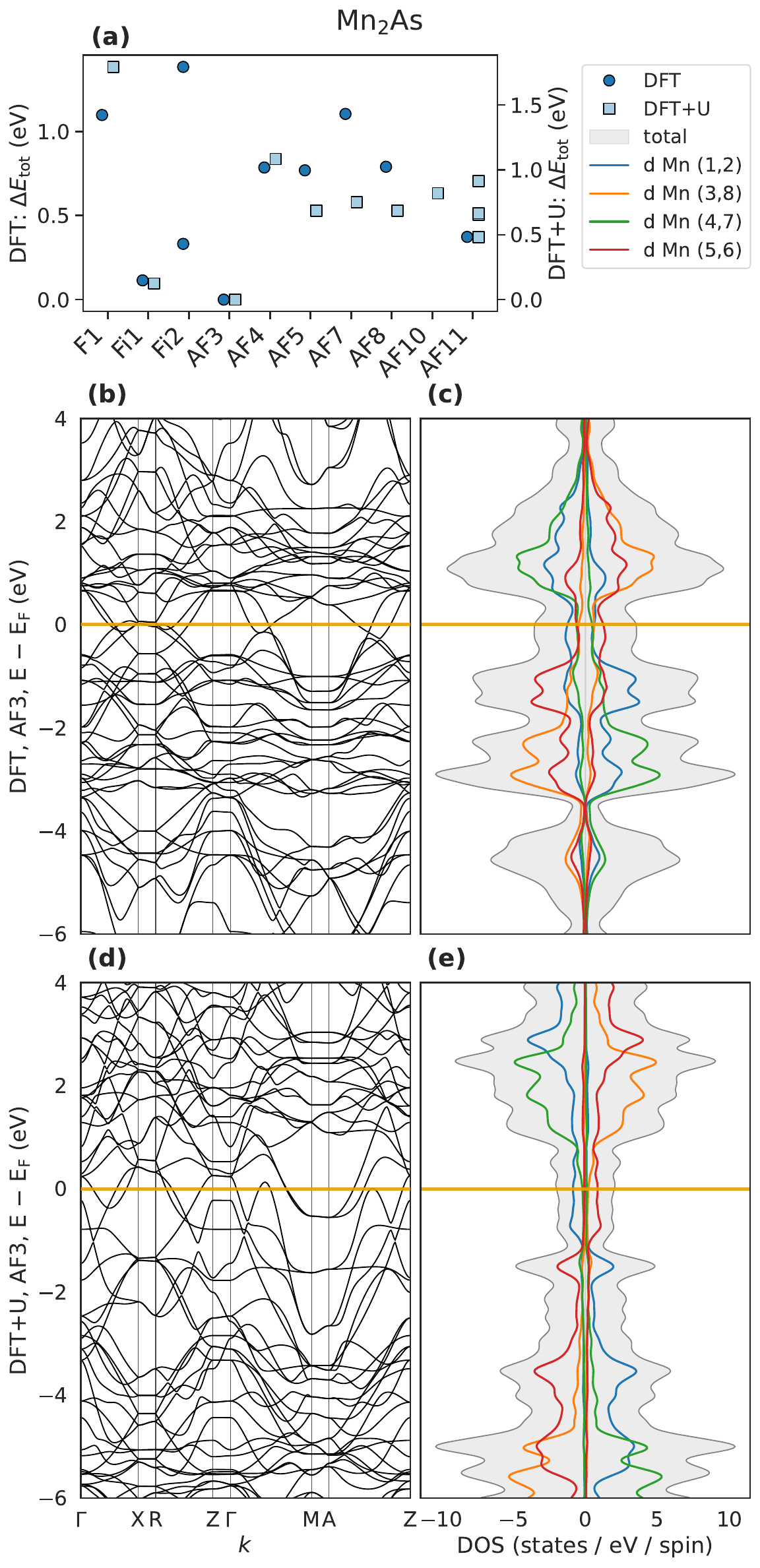}
\caption{(a) Optimized magnetic structures and their total energies relative to the lowest-energy configuration, shown for the DFT and DFT+$U$ methods. The electronic band structure and the $d$-orbital-projected density of states of the lowest-energy magnetic configuration obtained for Mn$_2$As using DFT are presented in panels (b) and (c), respectively. The corresponding DFT+$U$ results are shown in panels (d) and (e). For some atoms, the projected DOS curves are degenerate; the indices of these atoms are specified in the legend. The atomic indices correspond to those introduced in Fig.~\ref{fig:magnetic structures}.
}
\label{fig:dos bands}
\end{figure}

We first illustrate the workflow using Mn$_2$As as a representative example. The calculated lattice parameters are \(a=\qty{3.7517}{\angstrom}\) and \(c=\qty{6.1438}{\angstrom}\), while the independent fractional coordinates are \(z_\mathrm{Mn}=0.3350\) and \(z_\mathrm{As}=0.2645\). The calculated energies obtained using DFT and DFT+$U$ are shown in Fig.~\ref{fig:dos bands}(a). For both methods, the lowest-energy magnetic configuration is AF3, in agreement with experiment~\cite{austin1962}. Although the relative energies of some magnetic configurations differ significantly between DFT and DFT+$U$, both methods predict the same lowest-energy configuration. This behavior is also common among the other materials studied. The corresponding band structures and densities of states are shown in Figs.~\ref{fig:dos bands}(b) and \ref{fig:dos bands}(c) for plain DFT and in Figs.~\ref{fig:dos bands}(d) and \ref{fig:dos bands}(e) for DFT+$U$. The Supplementary Material contains the full catalog of calculated results for all investigated materials, covering the optimized lattice parameters and atomic positions, the converged magnetic structures and total energies, together with the density-of-states and band-structure plots corresponding to the lowest-energy calculations.

\begin{table}
\begin{tabular}{ll@{\hspace{2em}}ll@{\hspace{2em}}ll}
\toprule
Material & Label & Material & Label & Material & Label \\
\midrule
Co$_2$As & F1 & CrMnAs & AF5 & Ga$_2$As & NM \\
Co$_2$Sb & AF4 & CrMnSb & AF3 & Ga$_2$Sb & NM \\
CoCrAs & AF4 & CrNiSb & AF8'' & GaCrSb & AF2' \\
CoCuAs & NM & Cu$_2$As & NM & GaCuSb & NM \\
CoCuSb & NM & Cu$_2$Sb & NM & GaMnAs & AF2' \\
CoFeAs & AF4 & CuGaAs & NM & GaMnSb & AF2' \\
CoGaAs & F1'' & CuMnAs & AF2' & GaNiSb & NM \\
CoGaSb & F1'' & Fe$_2$As & AF4 & Mn$_2$As & AF3 \\
CoMnAs & AF4 & Fe$_2$Sb & AF4 & Mn$_2$Sb & Fi1 \\
CoMnSb & AF4 & FeCoSb & AF4 & MnCuSb & F1'' \\
CoNiAs & NM & FeCrAs & AF2' & Ni$_2$As & NM \\
CoNiSb & NM & FeCuAs & AF5'' & Ni$_2$Sb & NM \\
Cr$_2$As & AF5 & FeCuSb & F1'' & NiCrAs & F1' \\
Cr$_2$Sb & AF3 & FeGaAs & F1'' & NiCuAs & NM \\
CrCoSb & AF8'' & FeGaSb & AF5'' & NiCuSb & NM \\
CrCuAs & AF8'' & FeMnAs & AF3 & NiGaAs & NM \\
CrCuSb & AF5'' & FeMnSb & F1 & NiMnAs & AF3' \\
CrFeSb & AF11 & FeNiAs & AF1'' & NiMnSb & F1' \\
CrGaAs & AF8'' & FeNiSb & F1 &  &  \\
\bottomrule
\end{tabular}
\caption{For each material we present the lower-energy site-ordering variant. The lowest-energy magnetic configuration for each variant is shown in the label column. This is based on plain DFT calculations.}
\label{tab:materials and labels}
\end{table}

Comparison with the experimentally determined lattice parameters and atomic positions of known compounds, including Cr$_2$As~\cite{Pearson1985Cu2Sb}, Fe$_2$As~\cite{Nuss2006Fe2As}, Mn$_2$As~\cite{Nuss2006Fe2As}, Mn$_2$Sb~\cite{Pearson1985Cu2Sb}, CuMnAs~\cite{Wadley2015}, FeMnAs~\cite{Tobola2001}, Cu$_2$As~\cite{Naud1972}, Cu$_2$Sb~\cite{Nuss2002}, FeCuAs~\cite{Kamusella2017}, and FeCuSb~\cite{qian2012}, shows good overall agreement. The average relative differences are $\delta a=1.2\%$, $\delta c=1.6\%$, $\delta z_{M2}=2.0\%$, and $\delta z_{X}=1.5\%$, where $X$ denotes As or Sb. The largest deviations are found for the $z_{M2}$ coordinate in CuMnAs and Cr$_2$As, reaching approximately $4\%$.

In Tab.~\ref{tab:materials and labels} we show the lowest-energy magnetic structure for each calculated structure. Note that we label the structure analogously to CuMnAs, i.e., the first element occupies the M1 site, the second occupies the M2 site, and the final element is As or Sb. For the compounds containing two different transition metal elements, we calculate both possibilities (e.g. CuMnAs and MnCuAs) and include only the lowest-energy one here, although both are included in the Supplementary Material. We find that Mn and Cu are in vast majority of cases on the M2 site, whereas Co, Fe, and to a smaller extent also Cr, strongly prefer the M1 site. For Ga and Ni no preference is observed. It is not clear, however, how statistically sound these conclusions are, since the dataset we have is relatively small.

Comparison of the DFT-calculated lowest-energy magnetic structures with experimentally known magnetic structures shows agreement for all investigated materials: CuMnAs, Mn$_2$As~\cite{austin1962}, Fe$_2$As~\cite{katsuraki1966}, CuFeAs~\cite{thakur2014}, Mn$_2$Sb~\cite{wilkinson1957}, and CuFeSb~\cite{qian2012}, except Cr$_2$As~\cite{yamaguchi1972}. Experimental reports and previous theoretical studies identify the AF10 configuration as the magnetic ground state of Cr$_2$As, whereas our DFT calculations yield AF5 as the lowest-energy magnetic structure. The difference between AF5 and AF10 is $\qty{0.1}{\electronvolt}$, which is quite low compared to the range of the whole energy landscape $\qty{0.5}{\electronvolt}$ in the plain DFT calclations. This discrepancy in Cr$_2$As can be attributed to the structural parameters used in the calculations. The present results were obtained using lattice parameters and atomic positions calculated within DFT. In contrast, when the calculations were performed using the experimentally determined crystal structure of Cr$_2$As, the AF10 configuration was recovered as the lowest-energy magnetic state. Nevertheless, for the present workflow we consider the magnetic structure obtained from DFT-calculated lattice parameters and atomic positions to be more relevant, since experimentally determined structures are unavailable for many of the materials investigated here. It is therefore essential to assess the predictive capability of the workflow when applied to structural parameters obtained from first-principles calculations. The agreement with experiment obtained for all other materials provides confidence that calculations performed without experimental structural input can still yield physically realistic results. The DFT+$U$ calculations show agreement only for CuMnAs, Mn$_2$As and Mn$_2$Sb. Thus we expect in general that plain DFT calculations are more accurate. However, we provide the DFT+$U$ calculations for comparison, since in general correlations will be present that may not be captured by plain DFT calculations. 

\begin{figure}[ht]
\centering
\includegraphics[width=0.483\textwidth,page=1]{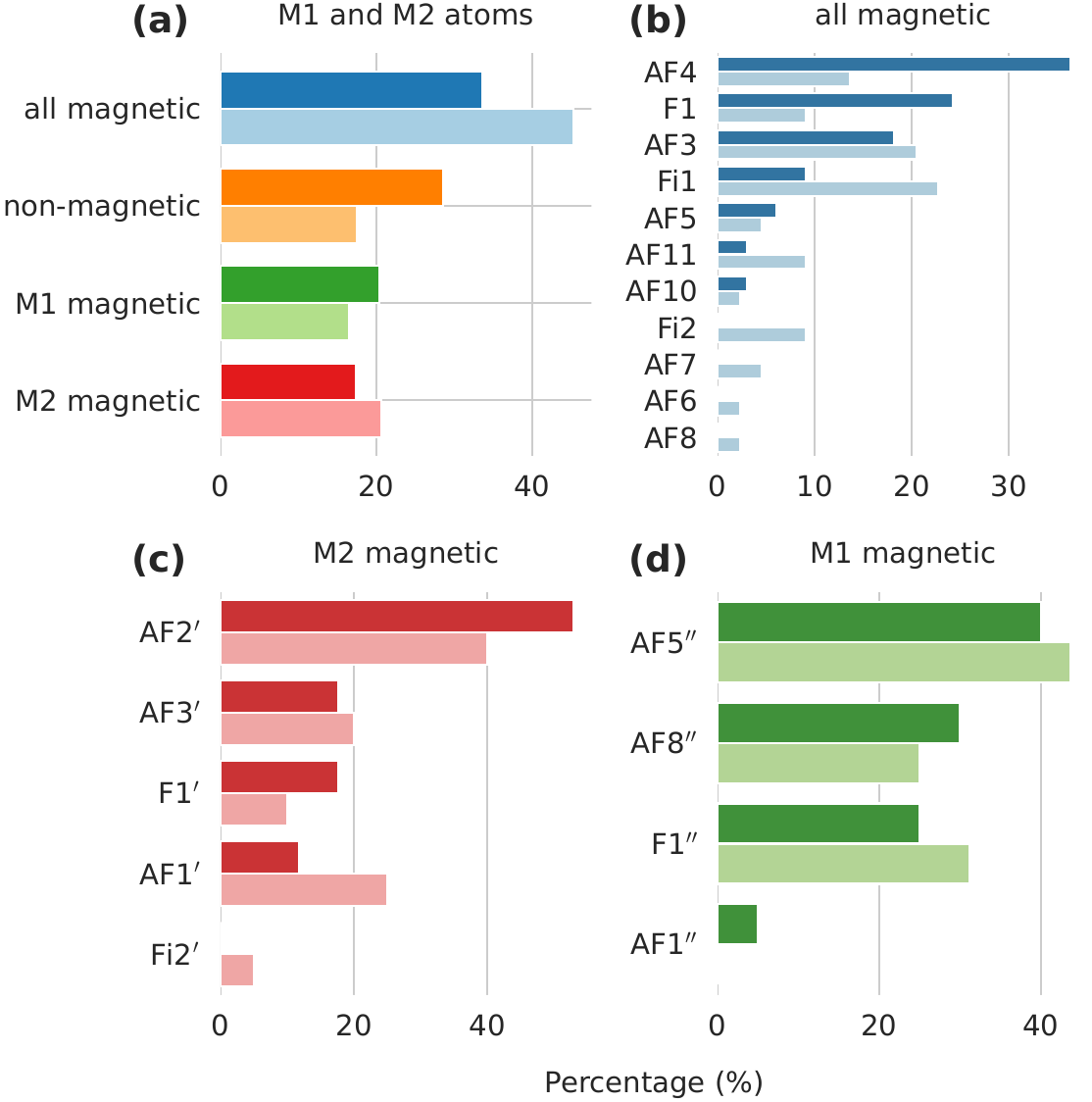}
\caption{Distribution of the energetically lowest magnetic configurations obtained for the investigated materials. (a) Histogram showing the percentage of magnetic structures in which both magnetic sublattices are magnetic (blue), only the M1 sublattice is magnetic (green), only the M2 sublattice is magnetic (red), or both sublattices are non-magnetic (orange). Panels (b)–(d) show the distributions of specific magnetic structures within the M1+M2, M2, and M1 categories, respectively.}
\label{fig:mag str distribution lowest}
\end{figure}

Having established the performance of the workflow for compounds with known
magnetic order, we now turn to the trends obtained across the full set of
calculated Cu$_2$Sb-type compounds. Here we consider both variants of the structure in the case where there are two different elements on the M1 and M2 sites. We first analyze the distribution of magnetic and non-magnetic ground states,
distinguishing between cases in which both M1 and M2 sites are magnetic,
only one site is magnetic, or both sites are non-magnetic
(Fig.~\ref{fig:mag str distribution lowest}(a)). Note that we consider a site non-magnetic if the magnetic moment is less than 0.1 $\mu_B$. We can conclude that the most populated state is the one where both M1 and M2 are magnetic. There are significant differences between DFT and DFT+$U$ calculations for both M1 and M2 magnetic and both M1 and M2 non-magnetic. As expected, DFT+$U$ increases the number of magnetic compounds. In Fig.~\ref{fig:mag str distribution lowest}(b) we show the distribution of lowest-energy magnetic configurations for the case where both M1 and M2 sites are magnetic. We find that the most common are two antiferromagnetic states, AF4 and AF3, as well as the ferromagnetic state F1 and the ferrimagnetic state Fi1. In contrast, the DFT+$U$ calculations most frequently favor the Fi1 configuration, followed by AF3. As shown in Fig.~\ref{fig:mag str distribution lowest}(c), in the case where only the M2 site is magnetic, the AF2$^{\prime}$ configuration, corresponding to the CuMnAs structure, is the most frequently obtained magnetic state for both DFT and DFT+$U$ calculations. For the case in which only the M1 sites are magnetic, presented in Fig.~\ref{fig:mag str distribution lowest}(d), the AF5$^{\prime\prime}$ structure occurs most frequently. Note that in some cases the DFT calculation converges to a state that is distinct from the magnetic configurations we consider. We use the label F2 if such a state is ferromagnetic, Fi2 if it is ferrimagnetic and AF11 if it is antiferromagnetic. The full converged magnetic configuration is given in the Supplementary Material.

\begin{figure*}[ht]
\centering
\includegraphics[width=1.0\textwidth,page=1]{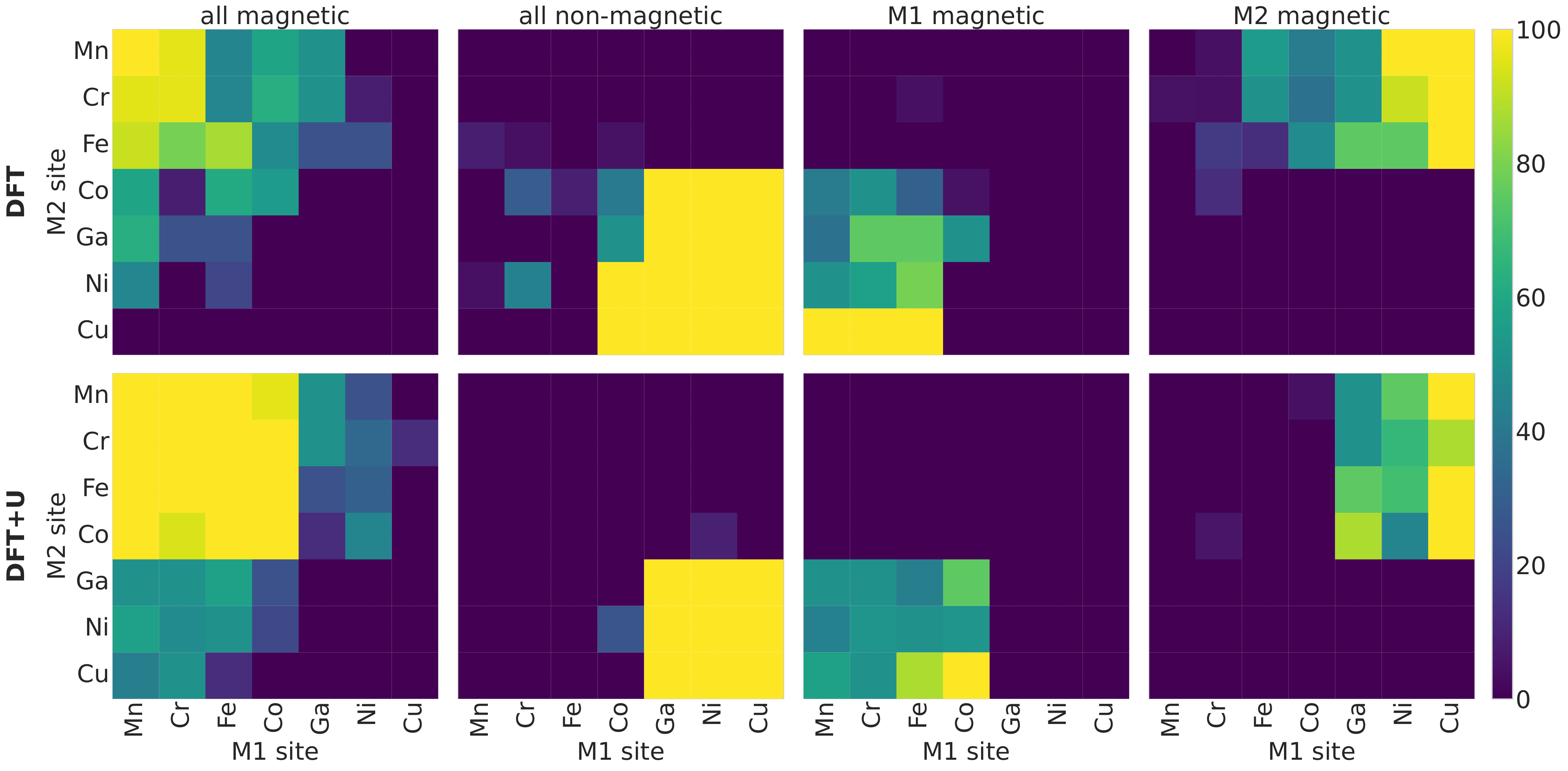}
\caption{For each combination of M1 and M2 atomic species, the color scale indicates the percentage of cases belonging to a given category: both M1 and M2 magnetic (first column), both M1 and M2 non-magnetic (second column), magnetic M1 and non-magnetic M2 (third column), and magnetic M2 and non-magnetic M1 (fourth column).}
\label{fig:magnetic-sites-heatmap}
\end{figure*}

\begin{figure}[ht]
\centering
\includegraphics[width=0.483\textwidth,page=1]{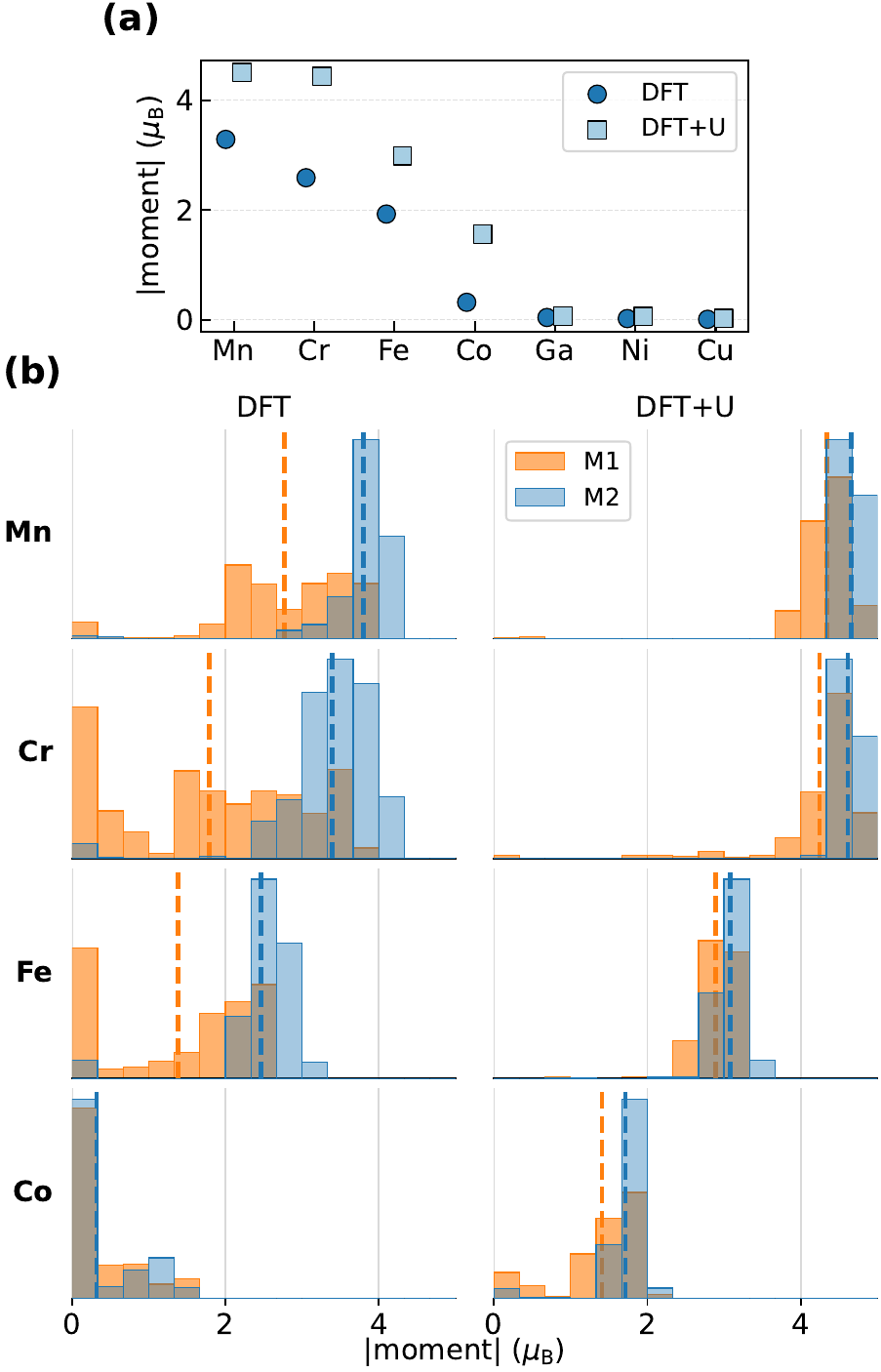}
\caption{(a) Average magnetic moment magnitude for each element; light and dark colors denote the DFT and DFT+$U$ methods, respectively.
(b) Element-resolved histograms of magnetic moments.}
\label{fig:magnetic moments}\end{figure}

Fig.~\ref{fig:magnetic-sites-heatmap} shows the distribution of the compounds into the four categories: both sites magnetic, both non-magnetic, only M1 magnetic and only M2 magnetic for each combination of elements occupying the M1 and M2 sites. 
Fig.~\ref{fig:magnetic moments}(a) shows the average magnetic moment magnitude for each element. Fig.~\ref{fig:magnetic moments}(b) shows the histogram of magnetic moments for specific elements on the M1 and M2 sites. Fig.~\ref{fig:magnetic-sites-heatmap} and Fig.~\ref{fig:magnetic moments} demonstrate that the formation and magnitude of local magnetic moments are governed by both the chemical identity of the element and its crystallographic site. Mn exhibits the most robust magnetic behavior, remaining magnetic on the M2 site in all cases and on the M1 site in nearly all cases. Cr and Fe also form stable moments on the M2 site, whereas their magnetic character on the M1 site is less robust and occurs only in approximately half of the investigated configurations. In contrast, Cu, Ni, and Ga exhibit only very small average moments and are never magnetic. This applies even for the DFT+$U$ calculations, however the DFT+$U$ increases the magnetism frequency and the magnetic moment magnitude. Furthermore, it reduces the distinction between the M1 and M2 sites. Co is an intermediate case, which is typically non-magnetic without U, but in DFT+$U$ calculation frequently becomes magnetic.

The Cu$_2$Sb compounds host a variety of magnetic orders with distinct symmetries, each of which may lend itself to a different application. In Tab.~\ref{tab:symmetries} we indicate, for a range of magnetic structures, whether inversion ($\mathcal{P}$), inversion combined with time-reversal ($\mathcal{PT}$), and translation combined with time-reversal ($\tau\mathcal{T}$) are preserved or broken. This analysis was performed using the Symmetr package~\cite{ZELEZNY2026}. We note that all of the antiferromagnetic orders possess either $\mathcal{PT}$ or $\tau\mathcal{T}$ symmetry, and thus none belong to the altermagnetic class~\cite{Smejkal2022}. Nevertheless, many effects of interest for spintronics can still be present in these compounds. In CuMnAs, spin-orbit torque has been used to electrically control the antiferromagnetic order~\cite{wadley2016}. This torque originates from the locally broken inversion symmetry at the magnetic sites. Owing to the symmetry of CuMnAs, it produces staggered effective magnetic fields---i.e.\ fields whose spatial pattern matches the antiferromagnetic order---which therefore couple to it efficiently. In the Cu$_2$Sb compounds, all magnetic sites locally break inversion symmetry, so a spin-orbit torque is present throughout the family. In some cases, however, the symmetry restricts this torque such that, in the strong-exchange limit, it does not couple efficiently to the magnetic order. For example, in the ferromagnetic case the inversion symmetry constrains the effective fields to be staggered, i.e.\ half of the sites experience fields of the opposite sign. Such fields would drive the system toward an antiferromagnetic state, which is prevented by the strong exchange. Analyzing the symmetry of all the magnetic orders, we find that, in general, the spin-orbit torque can couple efficiently to the magnetic order only if the global inversion symmetry is broken. This is because inversion connects sites with the same magnetic moment, whereas the effective fields on those sites must be opposite.
 
The symmetry also governs the character of the torque. In systems that break $\tau\mathcal{T}$ symmetry but possess $\mathcal{PT}$ symmetry, the effective fields are $\mathcal{T}$-even, typically resulting in a field-like torque, whereas in systems that possess $\tau\mathcal{T}$ symmetry but break $\mathcal{PT}$ symmetry, the effective fields are $\mathcal{T}$-odd and the torque is typically antidamping-like~\cite{Zelezny2017}. The $\mathcal{T}$-odd photocurrent that underlies the scanning technique of Ref.~\cite{schmid2026} can exist in systems that break both $\mathcal{P}$ and $\mathcal{PT}$ symmetry. In general, the existence of $\mathcal{T}$-odd effects is strongly restricted in $\tau\mathcal{T}$-symmetric systems, since any macroscopic effect is invariant under translations and must therefore be $\mathcal{T}$-even as a consequence of the $\tau\mathcal{T}$ symmetry. Only local effects, such as the spin-orbit torque, can be $\mathcal{T}$-odd in such systems. Note, however, that even these effects can give rise to macroscopic phenomena: the local torque can, for example, switch the magnetic order.
\begin{table}[h!]
\centering
\begin{tabular}{lccc}
\toprule
Structure & $\mathcal{P}$ & $\mathcal{PT}$ & $\tau\mathcal{T}$ \\
\midrule
F1 & \cmark & \xmark & \xmark \\
Fi1 & \cmark & \xmark & \xmark \\
AF1 & \xmark & \xmark & \cmark \\
AF2 & \xmark & \cmark & \xmark \\
AF3 & \cmark & \cmark & \cmark \\
AF4 & \cmark & \cmark & \cmark \\
AF5 & \xmark & \cmark & \xmark \\
AF6 & \xmark & \cmark & \xmark \\
AF7 & \xmark & \cmark & \xmark \\
AF8 & \xmark & \cmark & \xmark \\
AF9 & \xmark & \xmark & \cmark \\
AF10 & \cmark & \cmark & \cmark \\
AF1$^{\prime}$ & \cmark & \cmark & \cmark \\
AF2$^{\prime}$ & \xmark & \cmark & \xmark \\
AF3$^{\prime}$ & \cmark & \cmark & \cmark \\
AF1$^{\prime\prime}$ & \cmark & \cmark & \cmark \\
AF5$^{\prime\prime}$ & \xmark & \cmark & \xmark \\
AF8$^{\prime\prime}$ & \cmark & \cmark & \cmark \\
\bottomrule
\end{tabular}
\caption{Symmetries of all possible magnetic structures. The symmetries of F1$^{\prime}$ and F1$^{\prime\prime}$ are not shown, because they are the same as the symmetries of F1.}
\label{tab:symmetries}
\end{table}

\section{Conclusions}

Our work presents a comprehensive ab initio study of magnetic Cu$_2$Sb compounds, determining the lattice constants, magnetic order, and electronic structure for over 50 materials. Comparison with experimentally determined magnetic structures, where available, shows that our computational approach is accurate: plain DFT reproduces the measured ground state for all investigated compounds except Cr$_2$As, whose experimental order is recovered once the experimentally determined structure is used. DFT+$U$ agrees with experiment in fewer cases, and we therefore expect plain DFT to be the more reliable approach in general, while providing the DFT+$U$ results for comparison.
 
In agreement with previous results, we find the M2 site to be more commonly and more strongly magnetic than the M1 site. Mn is the most robustly magnetic: magnetic on M2 in every case and on M1 in nearly all. Cr and Fe are likewise always magnetic on M2, but on M1 they carry a moment in only about half of the configurations studied. Cu, Ni, and Ga are never magnetic; notably, Ni remains non-magnetic even in DFT+$U$, whereas Co is an intermediate case that is non-magnetic without U but frequently becomes magnetic once U is included.
 
The family hosts a variety of magnetic orders with distinct symmetries. Many may be promising for spintronics: a spin-orbit torque is present at every magnetic site, though our symmetry analysis shows it couples efficiently to the magnetic order only when global inversion symmetry is broken, and the character of the torque---field-like or antidamping-like---is set by whether $\mathcal{PT}$ or $\tau \mathcal{T}$ symmetry is preserved. Together with the possibility of engineering properties by combining elements or alloying, these results make the Cu$_2$Sb family a broad and largely unexplored platform, and our work is intended to serve as a basis for further theoretical and experimental studies.

\section{Methods} \label{sec:methods}

This section provides the technical details of the workflow summarized in Sec.~\ref{sec:workflow}. The basis functions used in OpenMX are pseudo-atomic orbitals (PAOs). These orbitals are pre-optimized within the OpenMX code~\cite{Ozaki2003}. Three predefined basis-set sizes are available: "Quick", "Standard", and "Precise". The energy cutoff parameter determines the numerical accuracy of the calculation of the difference-charge Coulomb potential, the exchange--correlation potential, and the solution of the Poisson equation performed using the fast Fourier transform.

The convergence of the computational parameters was tested with respect to the $k$-point mesh, PAO basis size, and energy cutoff. For each tested $k$-grid, a series of calculations was performed for CuMnAs with different magnetic configurations. The considered structures were the same as in Fig.~\ref{fig:magnetic structures}(c). The primary objective of the convergence testing was to verify that the energetic ordering of the magnetic configurations remained unchanged upon variation of the computational parameters. This condition was satisfied for all tested parameter values. Subsequently, a quantitative analysis of the parameter convergence was performed. Rather than comparing the absolute total energies, it is more meaningful to analyze the energy differences between magnetic structures. For the testing, the difference between the ferromagnetic (F1$^\prime{}$) and antiferromagnetic AF2$^\prime{}$ configurations was considered,

$$
\Delta E(x)=E_\mathrm{F1^{\prime}}(x)-E_{\mathrm{AF2}^\prime{}}(x),
$$

where $x$ denotes the tested computational parameter, such as the $k$-grid size, energy cutoff, or PAO basis size. The largest tested value of each parameter was used as a reference value $x_\mathrm{ref}$. The reference parameters were chosen as follows: $k$-grid $[48,48,48]$, PAO basis size "Precise", and energy cutoff $E_\mathrm{cut}=\qty{1000}{Ry}$. The difference in total energy relative to the reference structure was then quantified as

$$
\Delta U(x,x_\mathrm{ref})=\Delta E(x)-\Delta E(x_\mathrm{ref}).
$$

\begin{table}[]
\centering
\begin{tabular}{c|c}
$k$-grid & $\Delta U$ [meV] \\ \hline
6      & 40.8            \\
12     & -7.4            \\
24     & -1.3            
\end{tabular}
\caption{Dependence of the energy difference $\Delta U_i$ on the $k$-point grid $[i,i,i]$. The reference value, $\Delta E_{48}$, was obtained using the dense $k$-point grid $[48,48,48]$.}
\label{tab:kgrid}
\end{table}

The testing of the $k$-grid size was performed using the "Standard" PAO basis and an energy cutoff of $E_\mathrm{cut}=\qty{300}{Ry}$. A regular grid with the same number of k-points in every direction was used in this work. Tab.~\ref{tab:kgrid} summarizes the dependence of the energy difference $\Delta U_i$ on the $k$-grid density. The largest deviation was observed for the coarse mesh $[6,6,6]$, whereas the deviations corresponding to denser meshes were substantially smaller. Considering that the computational cost scales approximately cubically with the number of $k$-points along one reciprocal direction, the $k$-grid $[12,12,12]$ was selected as a suitable compromise between computational efficiency and numerical accuracy.

The difference between the "Quick" and "Precise" basis sets was found to be $\Delta U_{\mathrm{Quick}}=\qty{224.5}{\milli\electronvolt}$, while the difference between the "Standard" and "Precise" basis sets was $\Delta U_{\mathrm{Standard}}=\qty{58.5}{\milli\electronvolt}$. Since calculations performed with the "Precise" basis were approximately 2.5 times more computationally demanding than those using the "Standard" basis, the "Standard" setting was chosen for the subsequent calculations.

According to the OpenMX documentation, an energy cutoff of at least $\qty{300}{Ry}$ is recommended. In the present work, cutoff energies of $\qty{300}{Ry}$, $\qty{500}{Ry}$, and $\qty{1000}{Ry}$ were tested. The difference between calculations performed with $E_\mathrm{cut}=\qty{300}{Ry}$ and $E_\mathrm{cut}=\qty{1000}{Ry}$ was $\Delta U_{300}=\qty{11.3}{\milli\electronvolt}$, while the difference between $E_\mathrm{cut}=\qty{500}{Ry}$ and $E_\mathrm{cut}=\qty{1000}{Ry}$ was $\Delta U_{500}=\qty{4.5}{\milli\electronvolt}$. Based on these results, the value $E_\mathrm{cut}=\qty{300}{Ry}$ was considered sufficiently accurate for the purposes of the present study.

\subsection{Variable cell optimization (VCO)}

Several variable cell optimization (VCO) schemes are available in OpenMX, differing primarily in the symmetry constraints imposed during the structural optimization. In the present work, the RFC5 method~\cite{Ozaki2003} was employed, as it enables the simultaneous optimization of lattice parameters and internal atomic coordinates. The RFC5 optimization was performed in two consecutive steps. Since the method itself does not constrain the symmetry, small symmetry deviations may be present after the optimization. Thus, after the first converged calculation, the resulting structure was symmetrized to satisfy the crystallographic symmetry corresponding to the P4/nmm space group.

The symmetrized structure was subsequently used as the input for a second RFC5 optimization. After convergence, the resulting structure was again adjusted to satisfy the required P4/nmm symmetry. The initial input for the VCO calculations consisted of the experimentally determined lattice parameters and atomic positions of CuMnAs.

To test the accuracy of the VCO, we used CuMnAs, starting from a structure artificially distorted from the correct structure. Specifically, all fractional atomic coordinates were uniformly shifted by \(0.01\), while the lattice parameters \(a\) and \(b\) were increased by \(\qty{0.15}{\angstrom}\) and the lattice parameter \(c\) was decreased by \(\qty{0.4}{\angstrom}\).

We considered both spin-unpolarized and spin-polarized electronic configurations with the experimental magnetic structure of CuMnAs (AF2$^{\prime}$). The resulting optimized structures obtained from these two approaches differed substantially. The spin-unpolarized calculation showed a large difference from the experimental values: \(a_{\mathrm{rel}} = 8.1\%\) and \(c_{\mathrm{rel}} = 2.8\%\), while the relative differences in the fractional atomic $z$ coordinates were \(z^{\mathrm{Mn}}_{\mathrm{rel}} = 5.8\%\) and \(z^{\mathrm{As}}_{\mathrm{rel}} = 2.8\%\). The spin-polarized VCO calculation yielded significantly improved agreement with experiment. The relative differences in the lattice parameters were \(a_{\mathrm{rel}} = 2.3\%\) and \(c_{\mathrm{rel}} = 2.3\%\), whereas the relative differences in the atomic $z$ coordinates of Mn and As were \(z^{\mathrm{Mn}}_{\mathrm{rel}} = 4.0\%\) and \(z^{\mathrm{As}}_{\mathrm{rel}} = 1.0\%\), respectively.

These results demonstrate that spin-polarized VCO calculations reproduce the experimentally observed structure more accurately than spin-unpolarized calculations. A calculation using the ferromagnetic configuration produced a similar result. We therefore conclude that spin polarization is necessary for the VCO, whereas the specific magnetic configuration is less important. This conclusion is further supported by similar tests performed for Mn$_2$As.

Thus, we used a single magnetic structure for the VCO and then use the same optimized lattice for all magnetic configurations. For the spin-polarized VCO calculations, the AF5 magnetic configuration was employed when neither atomic species was Cu or Ga. For systems containing Cu or Ga, a different magnetic configuration was used instead: either AF2$^{\prime}$ or AF5$^{\prime\prime}$, depending on the position of the non-magnetic atom. All of these calculations were performed using a single-cell structure.

\subsection{Code Availability}

The OpenMX plugin that was developed for this project is published under open-source license and available on GitHub~\cite{aiida_openmx_plugin_2026}.

\section*{Author contributions}

JŽ concieved and supervised the project. VP carried out the calculations. Both authors have contributed to the development of the OpenMX plugin and to the manuscript preparation.

\vspace{1em}

\section*{Data Availability}

The data associated with the paper are included in the Supplemental Material or in the Zenodo repository. This includes the database of all the calculation results, and the relaxed structures.  

\begin{acknowledgments}
We acknowledge support from the Dioscuri Program LV23025 funded by MPG and MEYS, MEYS grant No. CZ.02.01.01/00/22\_008/0004594, GACR grant 25-18244S and ERC advanced grant 101095925. This work was supported by the Ministry of Education, Youth and Sports of the Czech Republic through the e-INFRA CZ (ID:90254).
\end{acknowledgments}

\bibliography{refs,bibliography} 

\begin{thebibliography}{39}%
\makeatletter
\providecommand \@ifxundefined [1]{%
 \@ifx{#1\undefined}
}%
\providecommand \@ifnum [1]{%
 \ifnum #1\expandafter \@firstoftwo
 \else \expandafter \@secondoftwo
 \fi
}%
\providecommand \@ifx [1]{%
 \ifx #1\expandafter \@firstoftwo
 \else \expandafter \@secondoftwo
 \fi
}%
\providecommand \natexlab [1]{#1}%
\providecommand \enquote  [1]{``#1''}%
\providecommand \bibnamefont  [1]{#1}%
\providecommand \bibfnamefont [1]{#1}%
\providecommand \citenamefont [1]{#1}%
\providecommand \href@noop [0]{\@secondoftwo}%
\providecommand \href [0]{\begingroup \@sanitize@url \@href}%
\providecommand \@href[1]{\@@startlink{#1}\@@href}%
\providecommand \@@href[1]{\endgroup#1\@@endlink}%
\providecommand \@sanitize@url [0]{\catcode `\\12\catcode `\$12\catcode
  `\&12\catcode `\#12\catcode `\^12\catcode `\_12\catcode `\%12\relax}%
\providecommand \@@startlink[1]{}%
\providecommand \@@endlink[0]{}%
\providecommand \url  [0]{\begingroup\@sanitize@url \@url }%
\providecommand \@url [1]{\endgroup\@href {#1}{\urlprefix }}%
\providecommand \urlprefix  [0]{URL }%
\providecommand \Eprint [0]{\href }%
\providecommand \doibase [0]{https://doi.org/}%
\providecommand \selectlanguage [0]{\@gobble}%
\providecommand \bibinfo  [0]{\@secondoftwo}%
\providecommand \bibfield  [0]{\@secondoftwo}%
\providecommand \translation [1]{[#1]}%
\providecommand \BibitemOpen [0]{}%
\providecommand \bibitemStop [0]{}%
\providecommand \bibitemNoStop [0]{.\EOS\space}%
\providecommand \EOS [0]{\spacefactor3000\relax}%
\providecommand \BibitemShut  [1]{\csname bibitem#1\endcsname}%
\let\auto@bib@innerbib\@empty
\bibitem [{\citenamefont {Wadley}\ \emph {et~al.}(2013)\citenamefont {Wadley},
  \citenamefont {Nov{\'a}k}, \citenamefont {Campion}, \citenamefont {Rinaldi},
  \citenamefont {Mart{\'i}}, \citenamefont {Reichlov{\'a}}, \citenamefont {{\v
  Z}elezn{\'y}}, \citenamefont {Gazquez}, \citenamefont {Roldan}, \citenamefont
  {Varela}, \citenamefont {Khalyavin}, \citenamefont {Langridge}, \citenamefont
  {Kriegner}, \citenamefont {M{\'a}ca}, \citenamefont {Ma{\v s}ek},
  \citenamefont {Bertacco}, \citenamefont {Hol{\'y}}, \citenamefont
  {Rushforth}, \citenamefont {Edmonds}, \citenamefont {Gallagher},
  \citenamefont {Foxon}, \citenamefont {Wunderlich},\ and\ \citenamefont
  {Jungwirth}}]{wadley2013}%
  \BibitemOpen
  \bibfield  {author} {\bibinfo {author} {\bibfnamefont {P.}~\bibnamefont
  {Wadley}}, \bibinfo {author} {\bibfnamefont {V.}~\bibnamefont {Nov{\'a}k}},
  \bibinfo {author} {\bibfnamefont {R.}~\bibnamefont {Campion}}, \bibinfo
  {author} {\bibfnamefont {C.}~\bibnamefont {Rinaldi}}, \bibinfo {author}
  {\bibfnamefont {X.}~\bibnamefont {Mart{\'i}}}, \bibinfo {author}
  {\bibfnamefont {H.}~\bibnamefont {Reichlov{\'a}}}, \bibinfo {author}
  {\bibfnamefont {J.}~\bibnamefont {{\v Z}elezn{\'y}}}, \bibinfo {author}
  {\bibfnamefont {J.}~\bibnamefont {Gazquez}}, \bibinfo {author} {\bibfnamefont
  {M.}~\bibnamefont {Roldan}}, \bibinfo {author} {\bibfnamefont
  {M.}~\bibnamefont {Varela}}, \bibinfo {author} {\bibfnamefont
  {D.}~\bibnamefont {Khalyavin}}, \bibinfo {author} {\bibfnamefont
  {S.}~\bibnamefont {Langridge}}, \bibinfo {author} {\bibfnamefont
  {D.}~\bibnamefont {Kriegner}}, \bibinfo {author} {\bibfnamefont
  {F.}~\bibnamefont {M{\'a}ca}}, \bibinfo {author} {\bibfnamefont
  {J.}~\bibnamefont {Ma{\v s}ek}}, \bibinfo {author} {\bibfnamefont
  {R.}~\bibnamefont {Bertacco}}, \bibinfo {author} {\bibfnamefont
  {V.}~\bibnamefont {Hol{\'y}}}, \bibinfo {author} {\bibfnamefont
  {A.}~\bibnamefont {Rushforth}}, \bibinfo {author} {\bibfnamefont
  {K.}~\bibnamefont {Edmonds}}, \bibinfo {author} {\bibfnamefont
  {B.}~\bibnamefont {Gallagher}}, \bibinfo {author} {\bibfnamefont
  {C.}~\bibnamefont {Foxon}}, \bibinfo {author} {\bibfnamefont
  {J.}~\bibnamefont {Wunderlich}},\ and\ \bibinfo {author} {\bibfnamefont
  {T.}~\bibnamefont {Jungwirth}},\ }\bibfield  {title} {\bibinfo {title}
  {Tetragonal phase of epitaxial room-temperature antiferromagnet {{CuMnAs}}},\
  }\href {https://doi.org/10.1038/ncomms3322} {\bibfield  {journal} {\bibinfo
  {journal} {Nature Communications}\ }\textbf {\bibinfo {volume} {4}},\
  \bibinfo {pages} {2322} (\bibinfo {year} {2013})}\BibitemShut {NoStop}%
\bibitem [{\citenamefont {Wadley}\ \emph {et~al.}(2016)\citenamefont {Wadley},
  \citenamefont {Howells}, \citenamefont {{\v Z}elezn{\'y}}, \citenamefont
  {Andrews}, \citenamefont {Hills}, \citenamefont {Campion}, \citenamefont
  {Nov{\'a}k}, \citenamefont {Olejn{\'i}k}, \citenamefont {Maccherozzi},
  \citenamefont {Dhesi}, \citenamefont {Martin}, \citenamefont {Wagner},
  \citenamefont {Wunderlich}, \citenamefont {Freimuth}, \citenamefont
  {Mokrousov}, \citenamefont {Kune{\v s}}, \citenamefont {Chauhan},
  \citenamefont {Grzybowski}, \citenamefont {Rushforth}, \citenamefont
  {Edmonds}, \citenamefont {Gallagher},\ and\ \citenamefont
  {Jungwirth}}]{wadley2016}%
  \BibitemOpen
  \bibfield  {author} {\bibinfo {author} {\bibfnamefont {P.}~\bibnamefont
  {Wadley}}, \bibinfo {author} {\bibfnamefont {B.}~\bibnamefont {Howells}},
  \bibinfo {author} {\bibfnamefont {J.}~\bibnamefont {{\v Z}elezn{\'y}}},
  \bibinfo {author} {\bibfnamefont {C.}~\bibnamefont {Andrews}}, \bibinfo
  {author} {\bibfnamefont {V.}~\bibnamefont {Hills}}, \bibinfo {author}
  {\bibfnamefont {R.~P.}\ \bibnamefont {Campion}}, \bibinfo {author}
  {\bibfnamefont {V.}~\bibnamefont {Nov{\'a}k}}, \bibinfo {author}
  {\bibfnamefont {K.}~\bibnamefont {Olejn{\'i}k}}, \bibinfo {author}
  {\bibfnamefont {F.}~\bibnamefont {Maccherozzi}}, \bibinfo {author}
  {\bibfnamefont {S.~S.}\ \bibnamefont {Dhesi}}, \bibinfo {author}
  {\bibfnamefont {S.~Y.}\ \bibnamefont {Martin}}, \bibinfo {author}
  {\bibfnamefont {T.}~\bibnamefont {Wagner}}, \bibinfo {author} {\bibfnamefont
  {J.}~\bibnamefont {Wunderlich}}, \bibinfo {author} {\bibfnamefont
  {F.}~\bibnamefont {Freimuth}}, \bibinfo {author} {\bibfnamefont
  {Y.}~\bibnamefont {Mokrousov}}, \bibinfo {author} {\bibfnamefont
  {J.}~\bibnamefont {Kune{\v s}}}, \bibinfo {author} {\bibfnamefont {J.~S.}\
  \bibnamefont {Chauhan}}, \bibinfo {author} {\bibfnamefont {M.~J.}\
  \bibnamefont {Grzybowski}}, \bibinfo {author} {\bibfnamefont {A.~W.}\
  \bibnamefont {Rushforth}}, \bibinfo {author} {\bibfnamefont {K.~W.}\
  \bibnamefont {Edmonds}}, \bibinfo {author} {\bibfnamefont {B.~L.}\
  \bibnamefont {Gallagher}},\ and\ \bibinfo {author} {\bibfnamefont
  {T.}~\bibnamefont {Jungwirth}},\ }\bibfield  {title} {\bibinfo {title}
  {Electrical switching of an antiferromagnet},\ }\href
  {https://doi.org/10.1126/science.aab1031} {\bibfield  {journal} {\bibinfo
  {journal} {Science}\ }\textbf {\bibinfo {volume} {351}},\ \bibinfo {pages}
  {587} (\bibinfo {year} {2016})}\BibitemShut {NoStop}%
\bibitem [{\citenamefont {{\v{Z}}elezn{\'{y}}}\ \emph
  {et~al.}(2017)\citenamefont {{\v{Z}}elezn{\'{y}}}, \citenamefont {Gao},
  \citenamefont {Manchon}, \citenamefont {Freimuth}, \citenamefont {Mokrousov},
  \citenamefont {Zemen}, \citenamefont {Ma{\v{s}}ek}, \citenamefont {Sinova},\
  and\ \citenamefont {Jungwirth}}]{Zelezny2017}%
  \BibitemOpen
  \bibfield  {author} {\bibinfo {author} {\bibfnamefont {J.}~\bibnamefont
  {{\v{Z}}elezn{\'{y}}}}, \bibinfo {author} {\bibfnamefont {H.}~\bibnamefont
  {Gao}}, \bibinfo {author} {\bibfnamefont {A.}~\bibnamefont {Manchon}},
  \bibinfo {author} {\bibfnamefont {F.}~\bibnamefont {Freimuth}}, \bibinfo
  {author} {\bibfnamefont {Y.}~\bibnamefont {Mokrousov}}, \bibinfo {author}
  {\bibfnamefont {J.}~\bibnamefont {Zemen}}, \bibinfo {author} {\bibfnamefont
  {J.}~\bibnamefont {Ma{\v{s}}ek}}, \bibinfo {author} {\bibfnamefont
  {J.}~\bibnamefont {Sinova}},\ and\ \bibinfo {author} {\bibfnamefont
  {T.}~\bibnamefont {Jungwirth}},\ }\bibfield  {title} {\bibinfo {title}
  {{Spin-orbit torques in locally and globally noncentrosymmetric crystals:
  Antiferromagnets and ferromagnets}},\ }\href
  {https://doi.org/10.1103/PhysRevB.95.014403} {\bibfield  {journal} {\bibinfo
  {journal} {Phys. Rev. B}\ }\textbf {\bibinfo {volume} {95}},\ \bibinfo
  {pages} {014403} (\bibinfo {year} {2017})}\BibitemShut {NoStop}%
\bibitem [{\citenamefont {Grzybowski}\ \emph {et~al.}(2017)\citenamefont
  {Grzybowski}, \citenamefont {Wadley}, \citenamefont {Edmonds}, \citenamefont
  {Beardsley}, \citenamefont {Hills}, \citenamefont {Campion}, \citenamefont
  {Gallagher}, \citenamefont {Chauhan}, \citenamefont {Novak}, \citenamefont
  {Jungwirth}, \citenamefont {Maccherozzi},\ and\ \citenamefont
  {Dhesi}}]{grzybowski2017}%
  \BibitemOpen
  \bibfield  {author} {\bibinfo {author} {\bibfnamefont {M.~J.}\ \bibnamefont
  {Grzybowski}}, \bibinfo {author} {\bibfnamefont {P.}~\bibnamefont {Wadley}},
  \bibinfo {author} {\bibfnamefont {K.~W.}\ \bibnamefont {Edmonds}}, \bibinfo
  {author} {\bibfnamefont {R.}~\bibnamefont {Beardsley}}, \bibinfo {author}
  {\bibfnamefont {V.}~\bibnamefont {Hills}}, \bibinfo {author} {\bibfnamefont
  {R.~P.}\ \bibnamefont {Campion}}, \bibinfo {author} {\bibfnamefont {B.~L.}\
  \bibnamefont {Gallagher}}, \bibinfo {author} {\bibfnamefont {J.~S.}\
  \bibnamefont {Chauhan}}, \bibinfo {author} {\bibfnamefont {V.}~\bibnamefont
  {Novak}}, \bibinfo {author} {\bibfnamefont {T.}~\bibnamefont {Jungwirth}},
  \bibinfo {author} {\bibfnamefont {F.}~\bibnamefont {Maccherozzi}},\ and\
  \bibinfo {author} {\bibfnamefont {S.~S.}\ \bibnamefont {Dhesi}},\ }\bibfield
  {title} {\bibinfo {title} {Imaging {{Current-Induced Switching}} of
  {{Antiferromagnetic Domains}} in {{CuMnAs}}},\ }\href
  {https://doi.org/10.1103/PhysRevLett.118.057701} {\bibfield  {journal}
  {\bibinfo  {journal} {Physical Review Letters}\ }\textbf {\bibinfo {volume}
  {118}},\ \bibinfo {pages} {057701} (\bibinfo {year} {2017})}\BibitemShut
  {NoStop}%
\bibitem [{\citenamefont {Olejn{\'i}k}\ \emph {et~al.}(2017)\citenamefont
  {Olejn{\'i}k}, \citenamefont {Schuler}, \citenamefont {Marti}, \citenamefont
  {Nov{\'a}k}, \citenamefont {Ka{\v s}par}, \citenamefont {Wadley},
  \citenamefont {Campion}, \citenamefont {Edmonds}, \citenamefont {Gallagher},
  \citenamefont {Garces}, \citenamefont {Baumgartner}, \citenamefont
  {Gambardella},\ and\ \citenamefont {Jungwirth}}]{olejnik2017}%
  \BibitemOpen
  \bibfield  {author} {\bibinfo {author} {\bibfnamefont {K.}~\bibnamefont
  {Olejn{\'i}k}}, \bibinfo {author} {\bibfnamefont {V.}~\bibnamefont
  {Schuler}}, \bibinfo {author} {\bibfnamefont {X.}~\bibnamefont {Marti}},
  \bibinfo {author} {\bibfnamefont {V.}~\bibnamefont {Nov{\'a}k}}, \bibinfo
  {author} {\bibfnamefont {Z.}~\bibnamefont {Ka{\v s}par}}, \bibinfo {author}
  {\bibfnamefont {P.}~\bibnamefont {Wadley}}, \bibinfo {author} {\bibfnamefont
  {R.~P.}\ \bibnamefont {Campion}}, \bibinfo {author} {\bibfnamefont {K.~W.}\
  \bibnamefont {Edmonds}}, \bibinfo {author} {\bibfnamefont {B.~L.}\
  \bibnamefont {Gallagher}}, \bibinfo {author} {\bibfnamefont {J.}~\bibnamefont
  {Garces}}, \bibinfo {author} {\bibfnamefont {M.}~\bibnamefont {Baumgartner}},
  \bibinfo {author} {\bibfnamefont {P.}~\bibnamefont {Gambardella}},\ and\
  \bibinfo {author} {\bibfnamefont {T.}~\bibnamefont {Jungwirth}},\ }\bibfield
  {title} {\bibinfo {title} {Antiferromagnetic {{CuMnAs}} multi-level memory
  cell with microelectronic compatibility},\ }\href
  {https://doi.org/10.1038/ncomms15434} {\bibfield  {journal} {\bibinfo
  {journal} {Nature Communications}\ }\textbf {\bibinfo {volume} {8}},\
  \bibinfo {pages} {15434} (\bibinfo {year} {2017})}\BibitemShut {NoStop}%
\bibitem [{\citenamefont {Olejn{\'i}k}\ \emph {et~al.}(2018)\citenamefont
  {Olejn{\'i}k}, \citenamefont {Seifert}, \citenamefont {Ka{\v s}par},
  \citenamefont {Nov{\'a}k}, \citenamefont {Wadley}, \citenamefont {Campion},
  \citenamefont {Baumgartner}, \citenamefont {Gambardella}, \citenamefont {N{\v
  e}mec}, \citenamefont {Wunderlich}, \citenamefont {Sinova}, \citenamefont
  {Ku{\v z}el}, \citenamefont {M{\"u}ller}, \citenamefont {Kampfrath},\ and\
  \citenamefont {Jungwirth}}]{olejnik2018}%
  \BibitemOpen
  \bibfield  {author} {\bibinfo {author} {\bibfnamefont {K.}~\bibnamefont
  {Olejn{\'i}k}}, \bibinfo {author} {\bibfnamefont {T.}~\bibnamefont
  {Seifert}}, \bibinfo {author} {\bibfnamefont {Z.}~\bibnamefont {Ka{\v
  s}par}}, \bibinfo {author} {\bibfnamefont {V.}~\bibnamefont {Nov{\'a}k}},
  \bibinfo {author} {\bibfnamefont {P.}~\bibnamefont {Wadley}}, \bibinfo
  {author} {\bibfnamefont {R.~P.}\ \bibnamefont {Campion}}, \bibinfo {author}
  {\bibfnamefont {M.}~\bibnamefont {Baumgartner}}, \bibinfo {author}
  {\bibfnamefont {P.}~\bibnamefont {Gambardella}}, \bibinfo {author}
  {\bibfnamefont {P.}~\bibnamefont {N{\v e}mec}}, \bibinfo {author}
  {\bibfnamefont {J.}~\bibnamefont {Wunderlich}}, \bibinfo {author}
  {\bibfnamefont {J.}~\bibnamefont {Sinova}}, \bibinfo {author} {\bibfnamefont
  {P.}~\bibnamefont {Ku{\v z}el}}, \bibinfo {author} {\bibfnamefont
  {M.}~\bibnamefont {M{\"u}ller}}, \bibinfo {author} {\bibfnamefont
  {T.}~\bibnamefont {Kampfrath}},\ and\ \bibinfo {author} {\bibfnamefont
  {T.}~\bibnamefont {Jungwirth}},\ }\bibfield  {title} {\bibinfo {title}
  {Terahertz electrical writing speed in an antiferromagnetic memory},\ }\href
  {https://doi.org/10.1126/sciadv.aar3566} {\bibfield  {journal} {\bibinfo
  {journal} {Science Advances}\ }\textbf {\bibinfo {volume} {4}},\ \bibinfo
  {pages} {eaar3566} (\bibinfo {year} {2018})}\BibitemShut {NoStop}%
\bibitem [{\citenamefont {Saidl}\ \emph {et~al.}(2017)\citenamefont {Saidl},
  \citenamefont {N{\v e}mec}, \citenamefont {Wadley}, \citenamefont {Hills},
  \citenamefont {Campion}, \citenamefont {Nov{\'a}k}, \citenamefont {Edmonds},
  \citenamefont {Maccherozzi}, \citenamefont {Dhesi}, \citenamefont
  {Gallagher}, \citenamefont {Troj{\'a}nek}, \citenamefont {Kune{\v s}},
  \citenamefont {{\v Z}elezn{\'y}}, \citenamefont {Mal{\'y}},\ and\
  \citenamefont {Jungwirth}}]{saidl2017}%
  \BibitemOpen
  \bibfield  {author} {\bibinfo {author} {\bibfnamefont {V.}~\bibnamefont
  {Saidl}}, \bibinfo {author} {\bibfnamefont {P.}~\bibnamefont {N{\v e}mec}},
  \bibinfo {author} {\bibfnamefont {P.}~\bibnamefont {Wadley}}, \bibinfo
  {author} {\bibfnamefont {V.}~\bibnamefont {Hills}}, \bibinfo {author}
  {\bibfnamefont {R.~P.}\ \bibnamefont {Campion}}, \bibinfo {author}
  {\bibfnamefont {V.}~\bibnamefont {Nov{\'a}k}}, \bibinfo {author}
  {\bibfnamefont {K.~W.}\ \bibnamefont {Edmonds}}, \bibinfo {author}
  {\bibfnamefont {F.}~\bibnamefont {Maccherozzi}}, \bibinfo {author}
  {\bibfnamefont {S.~S.}\ \bibnamefont {Dhesi}}, \bibinfo {author}
  {\bibfnamefont {B.~L.}\ \bibnamefont {Gallagher}}, \bibinfo {author}
  {\bibfnamefont {F.}~\bibnamefont {Troj{\'a}nek}}, \bibinfo {author}
  {\bibfnamefont {J.}~\bibnamefont {Kune{\v s}}}, \bibinfo {author}
  {\bibfnamefont {J.}~\bibnamefont {{\v Z}elezn{\'y}}}, \bibinfo {author}
  {\bibfnamefont {P.}~\bibnamefont {Mal{\'y}}},\ and\ \bibinfo {author}
  {\bibfnamefont {T.}~\bibnamefont {Jungwirth}},\ }\bibfield  {title} {\bibinfo
  {title} {Optical determination of the {{N\'eel}} vector in a {{CuMnAs}}
  thin-film antiferromagnet},\ }\href
  {https://doi.org/10.1038/nphoton.2016.255} {\bibfield  {journal} {\bibinfo
  {journal} {Nature Photonics}\ }\textbf {\bibinfo {volume} {11}},\ \bibinfo
  {pages} {91} (\bibinfo {year} {2017})}\BibitemShut {NoStop}%
\bibitem [{\citenamefont {Schmid}\ \emph {et~al.}(2026)\citenamefont {Schmid},
  \citenamefont {Siebenkotten}, \citenamefont {Dai}, \citenamefont {Godinho},
  \citenamefont {Ostatnick\'y}, \citenamefont {Zou}, \citenamefont {Zhang},
  \citenamefont {\v{Z}elezn\'y}, \citenamefont {\v{S}ob\'a\v{n}}, \citenamefont
  {K\v{r}\'i\v{z}ek}, \citenamefont {Nov\'ak}, \citenamefont {Fairman},
  \citenamefont {Hoehl}, \citenamefont {Hertwig}, \citenamefont {Janda},
  \citenamefont {Huber}, \citenamefont {Huber}, \citenamefont {K\"astner},\
  and\ \citenamefont {Wunderlich}}]{schmid2026}%
  \BibitemOpen
  \bibfield  {author} {\bibinfo {author} {\bibfnamefont {A.}~\bibnamefont
  {Schmid}}, \bibinfo {author} {\bibfnamefont {D.}~\bibnamefont
  {Siebenkotten}}, \bibinfo {author} {\bibfnamefont {D.}~\bibnamefont {Dai}},
  \bibinfo {author} {\bibfnamefont {J.}~\bibnamefont {Godinho}}, \bibinfo
  {author} {\bibfnamefont {T.}~\bibnamefont {Ostatnick\'y}}, \bibinfo {author}
  {\bibfnamefont {N.}~\bibnamefont {Zou}}, \bibinfo {author} {\bibfnamefont
  {Y.}~\bibnamefont {Zhang}}, \bibinfo {author} {\bibfnamefont
  {J.}~\bibnamefont {\v{Z}elezn\'y}}, \bibinfo {author} {\bibfnamefont
  {Z.}~\bibnamefont {\v{S}ob\'a\v{n}}}, \bibinfo {author} {\bibfnamefont
  {F.}~\bibnamefont {K\v{r}\'i\v{z}ek}}, \bibinfo {author} {\bibfnamefont
  {V.}~\bibnamefont {Nov\'ak}}, \bibinfo {author} {\bibfnamefont
  {S.}~\bibnamefont {Fairman}}, \bibinfo {author} {\bibfnamefont
  {A.}~\bibnamefont {Hoehl}}, \bibinfo {author} {\bibfnamefont
  {A.}~\bibnamefont {Hertwig}}, \bibinfo {author} {\bibfnamefont
  {T.}~\bibnamefont {Janda}}, \bibinfo {author} {\bibfnamefont {M.~A.}\
  \bibnamefont {Huber}}, \bibinfo {author} {\bibfnamefont {R.}~\bibnamefont
  {Huber}}, \bibinfo {author} {\bibfnamefont {B.}~\bibnamefont {K\"astner}},\
  and\ \bibinfo {author} {\bibfnamefont {J.}~\bibnamefont {Wunderlich}},\
  }\href {https://arxiv.org/abs/2604.21802} {\bibinfo {title} {Sign-resolved
  nanoscale readout and control of hidden antiferromagnetic spin order}}
  (\bibinfo {year} {2026}),\ \Eprint {https://arxiv.org/abs/2604.21802}
  {arXiv:2604.21802 [cond-mat.mtrl-sci]} \BibitemShut {NoStop}%
\bibitem [{\citenamefont {Krizek}\ \emph {et~al.}(2022)\citenamefont {Krizek},
  \citenamefont {Reimers}, \citenamefont {Ka{\v s}par}, \citenamefont
  {Marmodoro}, \citenamefont {Michali{\v c}ka}, \citenamefont {Man},
  \citenamefont {Edstr{\"o}m}, \citenamefont {Amin}, \citenamefont {Edmonds},
  \citenamefont {Campion}, \citenamefont {Maccherozzi}, \citenamefont {Dhesi},
  \citenamefont {Zub{\'a}{\v c}}, \citenamefont {Kriegner}, \citenamefont
  {Carbone}, \citenamefont {{\v Z}elezn{\'y}}, \citenamefont {V{\'y}born{\'y}},
  \citenamefont {Olejn{\'i}k}, \citenamefont {Nov{\'a}k}, \citenamefont {Rusz},
  \citenamefont {Idrobo}, \citenamefont {Wadley},\ and\ \citenamefont
  {Jungwirth}}]{krizek2022}%
  \BibitemOpen
  \bibfield  {author} {\bibinfo {author} {\bibfnamefont {F.}~\bibnamefont
  {Krizek}}, \bibinfo {author} {\bibfnamefont {S.}~\bibnamefont {Reimers}},
  \bibinfo {author} {\bibfnamefont {Z.}~\bibnamefont {Ka{\v s}par}}, \bibinfo
  {author} {\bibfnamefont {A.}~\bibnamefont {Marmodoro}}, \bibinfo {author}
  {\bibfnamefont {J.}~\bibnamefont {Michali{\v c}ka}}, \bibinfo {author}
  {\bibfnamefont {O.}~\bibnamefont {Man}}, \bibinfo {author} {\bibfnamefont
  {A.}~\bibnamefont {Edstr{\"o}m}}, \bibinfo {author} {\bibfnamefont {O.~J.}\
  \bibnamefont {Amin}}, \bibinfo {author} {\bibfnamefont {K.~W.}\ \bibnamefont
  {Edmonds}}, \bibinfo {author} {\bibfnamefont {R.~P.}\ \bibnamefont
  {Campion}}, \bibinfo {author} {\bibfnamefont {F.}~\bibnamefont
  {Maccherozzi}}, \bibinfo {author} {\bibfnamefont {S.~S.}\ \bibnamefont
  {Dhesi}}, \bibinfo {author} {\bibfnamefont {J.}~\bibnamefont {Zub{\'a}{\v
  c}}}, \bibinfo {author} {\bibfnamefont {D.}~\bibnamefont {Kriegner}},
  \bibinfo {author} {\bibfnamefont {D.}~\bibnamefont {Carbone}}, \bibinfo
  {author} {\bibfnamefont {J.}~\bibnamefont {{\v Z}elezn{\'y}}}, \bibinfo
  {author} {\bibfnamefont {K.}~\bibnamefont {V{\'y}born{\'y}}}, \bibinfo
  {author} {\bibfnamefont {K.}~\bibnamefont {Olejn{\'i}k}}, \bibinfo {author}
  {\bibfnamefont {V.}~\bibnamefont {Nov{\'a}k}}, \bibinfo {author}
  {\bibfnamefont {J.}~\bibnamefont {Rusz}}, \bibinfo {author} {\bibfnamefont
  {J.-C.}\ \bibnamefont {Idrobo}}, \bibinfo {author} {\bibfnamefont
  {P.}~\bibnamefont {Wadley}},\ and\ \bibinfo {author} {\bibfnamefont
  {T.}~\bibnamefont {Jungwirth}},\ }\bibfield  {title} {\bibinfo {title}
  {Atomically sharp domain walls in an antiferromagnet},\ }\href
  {https://doi.org/10.1126/sciadv.abn3535} {\bibfield  {journal} {\bibinfo
  {journal} {Science Advances}\ }\textbf {\bibinfo {volume} {8}},\ \bibinfo
  {pages} {eabn3535} (\bibinfo {year} {2022})}\BibitemShut {NoStop}%
\bibitem [{\citenamefont {Reimers}\ \emph {et~al.}(2024)\citenamefont
  {Reimers}, \citenamefont {Gomonay}, \citenamefont {Amin}, \citenamefont
  {Krizek}, \citenamefont {Barton}, \citenamefont {Lytvynenko}, \citenamefont
  {Poole}, \citenamefont {Nov{\'a}k}, \citenamefont {Campion}, \citenamefont
  {Maccherozzi}, \citenamefont {Carbone}, \citenamefont {Bj{\"o}rling},
  \citenamefont {Niu}, \citenamefont {Golias}, \citenamefont {Kriegner},
  \citenamefont {Sinova}, \citenamefont {Kl{\"a}ui}, \citenamefont {Jourdan},
  \citenamefont {Dhesi}, \citenamefont {Edmonds},\ and\ \citenamefont
  {Wadley}}]{reimers2024}%
  \BibitemOpen
  \bibfield  {author} {\bibinfo {author} {\bibfnamefont {S.}~\bibnamefont
  {Reimers}}, \bibinfo {author} {\bibfnamefont {O.}~\bibnamefont {Gomonay}},
  \bibinfo {author} {\bibfnamefont {O.~J.}\ \bibnamefont {Amin}}, \bibinfo
  {author} {\bibfnamefont {F.}~\bibnamefont {Krizek}}, \bibinfo {author}
  {\bibfnamefont {L.~X.}\ \bibnamefont {Barton}}, \bibinfo {author}
  {\bibfnamefont {Y.}~\bibnamefont {Lytvynenko}}, \bibinfo {author}
  {\bibfnamefont {S.~F.}\ \bibnamefont {Poole}}, \bibinfo {author}
  {\bibfnamefont {V.}~\bibnamefont {Nov{\'a}k}}, \bibinfo {author}
  {\bibfnamefont {R.~P.}\ \bibnamefont {Campion}}, \bibinfo {author}
  {\bibfnamefont {F.}~\bibnamefont {Maccherozzi}}, \bibinfo {author}
  {\bibfnamefont {G.}~\bibnamefont {Carbone}}, \bibinfo {author} {\bibfnamefont
  {A.}~\bibnamefont {Bj{\"o}rling}}, \bibinfo {author} {\bibfnamefont
  {Y.}~\bibnamefont {Niu}}, \bibinfo {author} {\bibfnamefont {E.}~\bibnamefont
  {Golias}}, \bibinfo {author} {\bibfnamefont {D.}~\bibnamefont {Kriegner}},
  \bibinfo {author} {\bibfnamefont {J.}~\bibnamefont {Sinova}}, \bibinfo
  {author} {\bibfnamefont {M.}~\bibnamefont {Kl{\"a}ui}}, \bibinfo {author}
  {\bibfnamefont {M.}~\bibnamefont {Jourdan}}, \bibinfo {author} {\bibfnamefont
  {S.~S.}\ \bibnamefont {Dhesi}}, \bibinfo {author} {\bibfnamefont {K.~W.}\
  \bibnamefont {Edmonds}},\ and\ \bibinfo {author} {\bibfnamefont
  {P.}~\bibnamefont {Wadley}},\ }\bibfield  {title} {\bibinfo {title} {Magnetic
  domain engineering in antiferromagnetic {{Cu Mn As}} and {{Mn}} 2 {{Au}}},\
  }\href {https://doi.org/10.1103/PhysRevApplied.21.064030} {\bibfield
  {journal} {\bibinfo  {journal} {Physical Review Applied}\ }\textbf {\bibinfo
  {volume} {21}},\ \bibinfo {pages} {064030} (\bibinfo {year}
  {2024})}\BibitemShut {NoStop}%
\bibitem [{\citenamefont {Grzybowski}\ \emph {et~al.}(2019)\citenamefont
  {Grzybowski}, \citenamefont {Wadley}, \citenamefont {Edmonds}, \citenamefont
  {Campion}, \citenamefont {Dybko}, \citenamefont {Majewicz}, \citenamefont
  {Gallagher}, \citenamefont {Sawicki},\ and\ \citenamefont
  {Dietl}}]{grzybowski2019}%
  \BibitemOpen
  \bibfield  {author} {\bibinfo {author} {\bibfnamefont {M.~J.}\ \bibnamefont
  {Grzybowski}}, \bibinfo {author} {\bibfnamefont {P.}~\bibnamefont {Wadley}},
  \bibinfo {author} {\bibfnamefont {K.~W.}\ \bibnamefont {Edmonds}}, \bibinfo
  {author} {\bibfnamefont {R.~P.}\ \bibnamefont {Campion}}, \bibinfo {author}
  {\bibfnamefont {K.}~\bibnamefont {Dybko}}, \bibinfo {author} {\bibfnamefont
  {M.}~\bibnamefont {Majewicz}}, \bibinfo {author} {\bibfnamefont {B.~L.}\
  \bibnamefont {Gallagher}}, \bibinfo {author} {\bibfnamefont {M.}~\bibnamefont
  {Sawicki}},\ and\ \bibinfo {author} {\bibfnamefont {T.}~\bibnamefont
  {Dietl}},\ }\bibfield  {title} {\bibinfo {title} {Gating effects in
  antiferromagnetic {{CuMnAs}}},\ }\href {https://doi.org/10.1063/1.5124354}
  {\bibfield  {journal} {\bibinfo  {journal} {AIP Advances}\ }\textbf {\bibinfo
  {volume} {9}},\ \bibinfo {pages} {115101} (\bibinfo {year}
  {2019})}\BibitemShut {NoStop}%
\bibitem [{\citenamefont {Godinho}\ \emph {et~al.}(2018)\citenamefont
  {Godinho}, \citenamefont {Reichlov{\'a}}, \citenamefont {Kriegner},
  \citenamefont {Nov{\'a}k}, \citenamefont {Olejn{\'i}k}, \citenamefont {Ka{\v
  s}par}, \citenamefont {{\v S}ob{\'a}{\v n}}, \citenamefont {Wadley},
  \citenamefont {Campion}, \citenamefont {Otxoa}, \citenamefont {Roy},
  \citenamefont {{\v Z}elezn{\'y}}, \citenamefont {Jungwirth},\ and\
  \citenamefont {Wunderlich}}]{godinho2018}%
  \BibitemOpen
  \bibfield  {author} {\bibinfo {author} {\bibfnamefont {J.}~\bibnamefont
  {Godinho}}, \bibinfo {author} {\bibfnamefont {H.}~\bibnamefont
  {Reichlov{\'a}}}, \bibinfo {author} {\bibfnamefont {D.}~\bibnamefont
  {Kriegner}}, \bibinfo {author} {\bibfnamefont {V.}~\bibnamefont {Nov{\'a}k}},
  \bibinfo {author} {\bibfnamefont {K.}~\bibnamefont {Olejn{\'i}k}}, \bibinfo
  {author} {\bibfnamefont {Z.}~\bibnamefont {Ka{\v s}par}}, \bibinfo {author}
  {\bibfnamefont {Z.}~\bibnamefont {{\v S}ob{\'a}{\v n}}}, \bibinfo {author}
  {\bibfnamefont {P.}~\bibnamefont {Wadley}}, \bibinfo {author} {\bibfnamefont
  {R.~P.}\ \bibnamefont {Campion}}, \bibinfo {author} {\bibfnamefont {R.~M.}\
  \bibnamefont {Otxoa}}, \bibinfo {author} {\bibfnamefont {P.~E.}\ \bibnamefont
  {Roy}}, \bibinfo {author} {\bibfnamefont {J.}~\bibnamefont {{\v
  Z}elezn{\'y}}}, \bibinfo {author} {\bibfnamefont {T.}~\bibnamefont
  {Jungwirth}},\ and\ \bibinfo {author} {\bibfnamefont {J.}~\bibnamefont
  {Wunderlich}},\ }\bibfield  {title} {\bibinfo {title} {Electrically induced
  and detected {{N\'eel}} vector reversal in a collinear antiferromagnet},\
  }\href {https://doi.org/10.1038/s41467-018-07092-2} {\bibfield  {journal}
  {\bibinfo  {journal} {Nature Communications}\ }\textbf {\bibinfo {volume}
  {9}},\ \bibinfo {pages} {4686} (\bibinfo {year} {2018})}\BibitemShut
  {NoStop}%
\bibitem [{\citenamefont {Ka{\v s}par}\ \emph {et~al.}(2020)\citenamefont
  {Ka{\v s}par}, \citenamefont {Sur{\'y}nek}, \citenamefont {Zub{\'a}{\v c}},
  \citenamefont {Krizek}, \citenamefont {Nov{\'a}k}, \citenamefont {Campion},
  \citenamefont {W{\"o}rnle}, \citenamefont {Gambardella}, \citenamefont
  {Marti}, \citenamefont {N{\v e}mec}, \citenamefont {Edmonds}, \citenamefont
  {Reimers}, \citenamefont {Amin}, \citenamefont {Maccherozzi}, \citenamefont
  {Dhesi}, \citenamefont {Wadley}, \citenamefont {Wunderlich}, \citenamefont
  {Olejn{\'i}k},\ and\ \citenamefont {Jungwirth}}]{kaspar2020}%
  \BibitemOpen
  \bibfield  {author} {\bibinfo {author} {\bibfnamefont {Z.}~\bibnamefont
  {Ka{\v s}par}}, \bibinfo {author} {\bibfnamefont {M.}~\bibnamefont
  {Sur{\'y}nek}}, \bibinfo {author} {\bibfnamefont {J.}~\bibnamefont
  {Zub{\'a}{\v c}}}, \bibinfo {author} {\bibfnamefont {F.}~\bibnamefont
  {Krizek}}, \bibinfo {author} {\bibfnamefont {V.}~\bibnamefont {Nov{\'a}k}},
  \bibinfo {author} {\bibfnamefont {R.~P.}\ \bibnamefont {Campion}}, \bibinfo
  {author} {\bibfnamefont {M.~S.}\ \bibnamefont {W{\"o}rnle}}, \bibinfo
  {author} {\bibfnamefont {P.}~\bibnamefont {Gambardella}}, \bibinfo {author}
  {\bibfnamefont {X.}~\bibnamefont {Marti}}, \bibinfo {author} {\bibfnamefont
  {P.}~\bibnamefont {N{\v e}mec}}, \bibinfo {author} {\bibfnamefont {K.~W.}\
  \bibnamefont {Edmonds}}, \bibinfo {author} {\bibfnamefont {S.}~\bibnamefont
  {Reimers}}, \bibinfo {author} {\bibfnamefont {O.~J.}\ \bibnamefont {Amin}},
  \bibinfo {author} {\bibfnamefont {F.}~\bibnamefont {Maccherozzi}}, \bibinfo
  {author} {\bibfnamefont {S.~S.}\ \bibnamefont {Dhesi}}, \bibinfo {author}
  {\bibfnamefont {P.}~\bibnamefont {Wadley}}, \bibinfo {author} {\bibfnamefont
  {J.}~\bibnamefont {Wunderlich}}, \bibinfo {author} {\bibfnamefont
  {K.}~\bibnamefont {Olejn{\'i}k}},\ and\ \bibinfo {author} {\bibfnamefont
  {T.}~\bibnamefont {Jungwirth}},\ }\bibfield  {title} {\bibinfo {title}
  {Quenching of an antiferromagnet into high resistivity states using
  electrical or ultrashort optical pulses},\ }\href
  {https://doi.org/10.1038/s41928-020-00506-4} {\bibfield  {journal} {\bibinfo
  {journal} {Nature Electronics}\ }\textbf {\bibinfo {volume} {4}},\ \bibinfo
  {pages} {30} (\bibinfo {year} {2020})}\BibitemShut {NoStop}%
\bibitem [{\citenamefont {Zub{\'a}{\v c}}\ \emph {et~al.}(2021)\citenamefont
  {Zub{\'a}{\v c}}, \citenamefont {Ka{\v s}par}, \citenamefont {Krizek},
  \citenamefont {F{\"o}rster}, \citenamefont {Campion}, \citenamefont
  {Nov{\'a}k}, \citenamefont {Jungwirth},\ and\ \citenamefont
  {Olejn{\'i}k}}]{zubac2021}%
  \BibitemOpen
  \bibfield  {author} {\bibinfo {author} {\bibfnamefont {J.}~\bibnamefont
  {Zub{\'a}{\v c}}}, \bibinfo {author} {\bibfnamefont {Z.}~\bibnamefont {Ka{\v
  s}par}}, \bibinfo {author} {\bibfnamefont {F.}~\bibnamefont {Krizek}},
  \bibinfo {author} {\bibfnamefont {T.}~\bibnamefont {F{\"o}rster}}, \bibinfo
  {author} {\bibfnamefont {R.~P.}\ \bibnamefont {Campion}}, \bibinfo {author}
  {\bibfnamefont {V.}~\bibnamefont {Nov{\'a}k}}, \bibinfo {author}
  {\bibfnamefont {T.}~\bibnamefont {Jungwirth}},\ and\ \bibinfo {author}
  {\bibfnamefont {K.}~\bibnamefont {Olejn{\'i}k}},\ }\bibfield  {title}
  {\bibinfo {title} {Hysteretic effects and magnetotransport of electrically
  switched {{CuMnAs}}},\ }\href {https://doi.org/10.1103/PhysRevB.104.184424}
  {\bibfield  {journal} {\bibinfo  {journal} {Physical Review B}\ }\textbf
  {\bibinfo {volume} {104}},\ \bibinfo {pages} {184424} (\bibinfo {year}
  {2021})}\BibitemShut {NoStop}%
\bibitem [{\citenamefont {Sur{\'y}nek}\ \emph {et~al.}(2025)\citenamefont
  {Sur{\'y}nek}, \citenamefont {Zub{\'a}{\v c}}, \citenamefont {Olejn{\'i}k},
  \citenamefont {Farka{\v s}}, \citenamefont {K{\v r}{\'i}{\v z}ek},
  \citenamefont {N{\'a}dvorn{\'i}k}, \citenamefont {Kuba{\v s}{\v c}{\'i}k},
  \citenamefont {Troj{\'a}nek}, \citenamefont {Campion}, \citenamefont
  {Nov{\'a}k}, \citenamefont {Jungwirth},\ and\ \citenamefont {N{\v
  e}mec}}]{surynek2025}%
  \BibitemOpen
  \bibfield  {author} {\bibinfo {author} {\bibfnamefont {M.}~\bibnamefont
  {Sur{\'y}nek}}, \bibinfo {author} {\bibfnamefont {J.}~\bibnamefont
  {Zub{\'a}{\v c}}}, \bibinfo {author} {\bibfnamefont {K.}~\bibnamefont
  {Olejn{\'i}k}}, \bibinfo {author} {\bibfnamefont {A.}~\bibnamefont {Farka{\v
  s}}}, \bibinfo {author} {\bibfnamefont {F.}~\bibnamefont {K{\v r}{\'i}{\v
  z}ek}}, \bibinfo {author} {\bibfnamefont {L.}~\bibnamefont
  {N{\'a}dvorn{\'i}k}}, \bibinfo {author} {\bibfnamefont {P.}~\bibnamefont
  {Kuba{\v s}{\v c}{\'i}k}}, \bibinfo {author} {\bibfnamefont {F.}~\bibnamefont
  {Troj{\'a}nek}}, \bibinfo {author} {\bibfnamefont {R.~P.}\ \bibnamefont
  {Campion}}, \bibinfo {author} {\bibfnamefont {V.}~\bibnamefont {Nov{\'a}k}},
  \bibinfo {author} {\bibfnamefont {T.}~\bibnamefont {Jungwirth}},\ and\
  \bibinfo {author} {\bibfnamefont {P.}~\bibnamefont {N{\v e}mec}},\ }\bibfield
   {title} {\bibinfo {title} {Picosecond transfer from short-term to long-term
  memory in analog antiferromagnetic memory device},\ }\href
  {https://doi.org/10.1016/j.newton.2025.100034} {\bibfield  {journal}
  {\bibinfo  {journal} {Newton}\ }\textbf {\bibinfo {volume} {1}},\ \bibinfo
  {pages} {100034} (\bibinfo {year} {2025})}\BibitemShut {NoStop}%
\bibitem [{\citenamefont {Zub{\'a}{\v c}}\ \emph {et~al.}(2025)\citenamefont
  {Zub{\'a}{\v c}}, \citenamefont {Sur{\'y}nek}, \citenamefont {Olejn{\'i}k},
  \citenamefont {Farka{\v s}}, \citenamefont {Krizek}, \citenamefont
  {N{\'a}dvorn{\'i}k}, \citenamefont {Kuba{\v s}{\v c}{\'i}k}, \citenamefont
  {Ka{\v s}par}, \citenamefont {Troj{\'a}nek}, \citenamefont {Campion},
  \citenamefont {Nov{\'a}k}, \citenamefont {N{\v e}mec},\ and\ \citenamefont
  {Jungwirth}}]{zubac2025}%
  \BibitemOpen
  \bibfield  {author} {\bibinfo {author} {\bibfnamefont {J.}~\bibnamefont
  {Zub{\'a}{\v c}}}, \bibinfo {author} {\bibfnamefont {M.}~\bibnamefont
  {Sur{\'y}nek}}, \bibinfo {author} {\bibfnamefont {K.}~\bibnamefont
  {Olejn{\'i}k}}, \bibinfo {author} {\bibfnamefont {A.}~\bibnamefont {Farka{\v
  s}}}, \bibinfo {author} {\bibfnamefont {F.}~\bibnamefont {Krizek}}, \bibinfo
  {author} {\bibfnamefont {L.}~\bibnamefont {N{\'a}dvorn{\'i}k}}, \bibinfo
  {author} {\bibfnamefont {P.}~\bibnamefont {Kuba{\v s}{\v c}{\'i}k}}, \bibinfo
  {author} {\bibfnamefont {Z.}~\bibnamefont {Ka{\v s}par}}, \bibinfo {author}
  {\bibfnamefont {F.}~\bibnamefont {Troj{\'a}nek}}, \bibinfo {author}
  {\bibfnamefont {R.~P.}\ \bibnamefont {Campion}}, \bibinfo {author}
  {\bibfnamefont {V.}~\bibnamefont {Nov{\'a}k}}, \bibinfo {author}
  {\bibfnamefont {P.}~\bibnamefont {N{\v e}mec}},\ and\ \bibinfo {author}
  {\bibfnamefont {T.}~\bibnamefont {Jungwirth}},\ }\bibfield  {title} {\bibinfo
  {title} {Investigation of {{Opto}}-magnetic {{Memory Effects}} in
  {{Antiferromagnetic CuMnAs Using Ultrafast Heat Dynamics}} and {{Quench
  Switching}}},\ }\href {https://doi.org/10.1002/aelm.202400835} {\bibfield
  {journal} {\bibinfo  {journal} {Advanced Electronic Materials}\ }\textbf
  {\bibinfo {volume} {11}},\ \bibinfo {pages} {2400835} (\bibinfo {year}
  {2025})}\BibitemShut {NoStop}%
\bibitem [{\citenamefont {Olejn{\'i}k}\ \emph {et~al.}(2025)\citenamefont
  {Olejn{\'i}k}, \citenamefont {Ka{\v s}par}, \citenamefont {Zub{\'a}{\v c}},
  \citenamefont {Telkamp}, \citenamefont {Farka{\v s}}, \citenamefont
  {Kriegner}, \citenamefont {V{\'y}born{\'y}}, \citenamefont {{\v
  Z}elezn{\'y}}, \citenamefont {{\v S}ob{\'a}{\v n}}, \citenamefont {Zeng},
  \citenamefont {Jungwirth}, \citenamefont {Nov{\'a}k},\ and\ \citenamefont
  {Krizek}}]{olejnik2025}%
  \BibitemOpen
  \bibfield  {author} {\bibinfo {author} {\bibfnamefont {K.}~\bibnamefont
  {Olejn{\'i}k}}, \bibinfo {author} {\bibfnamefont {Z.}~\bibnamefont {Ka{\v
  s}par}}, \bibinfo {author} {\bibfnamefont {J.}~\bibnamefont {Zub{\'a}{\v
  c}}}, \bibinfo {author} {\bibfnamefont {S.}~\bibnamefont {Telkamp}}, \bibinfo
  {author} {\bibfnamefont {A.}~\bibnamefont {Farka{\v s}}}, \bibinfo {author}
  {\bibfnamefont {D.}~\bibnamefont {Kriegner}}, \bibinfo {author}
  {\bibfnamefont {K.}~\bibnamefont {V{\'y}born{\'y}}}, \bibinfo {author}
  {\bibfnamefont {J.}~\bibnamefont {{\v Z}elezn{\'y}}}, \bibinfo {author}
  {\bibfnamefont {Z.}~\bibnamefont {{\v S}ob{\'a}{\v n}}}, \bibinfo {author}
  {\bibfnamefont {P.}~\bibnamefont {Zeng}}, \bibinfo {author} {\bibfnamefont
  {T.}~\bibnamefont {Jungwirth}}, \bibinfo {author} {\bibfnamefont
  {V.}~\bibnamefont {Nov{\'a}k}},\ and\ \bibinfo {author} {\bibfnamefont
  {F.}~\bibnamefont {Krizek}},\ }\bibfield  {title} {\bibinfo {title} {Quench
  switching of {{Mn}} 2 {{As}}},\ }\href {https://doi.org/10.1103/1c81-tnzw}
  {\bibfield  {journal} {\bibinfo  {journal} {Physical Review B}\ }\textbf
  {\bibinfo {volume} {112}},\ \bibinfo {pages} {144401} (\bibinfo {year}
  {2025})}\BibitemShut {NoStop}%
\bibitem [{\citenamefont {Austin}\ \emph {et~al.}(1962)\citenamefont {Austin},
  \citenamefont {Adelson},\ and\ \citenamefont {Cloud}}]{austin1962}%
  \BibitemOpen
  \bibfield  {author} {\bibinfo {author} {\bibfnamefont {A.~E.}\ \bibnamefont
  {Austin}}, \bibinfo {author} {\bibfnamefont {E.}~\bibnamefont {Adelson}},\
  and\ \bibinfo {author} {\bibfnamefont {W.~H.}\ \bibnamefont {Cloud}},\
  }\bibfield  {title} {\bibinfo {title} {Magnetic {{Structures}} of {{Mn2As}}
  and {{Mn2Sb0}}.{{7As0}}.3},\ }\href {https://doi.org/10.1063/1.1728729}
  {\bibfield  {journal} {\bibinfo  {journal} {Journal of Applied Physics}\
  }\textbf {\bibinfo {volume} {33}},\ \bibinfo {pages} {1356} (\bibinfo {year}
  {1962})}\BibitemShut {NoStop}%
\bibitem [{\citenamefont {Yamaguchi}\ \emph {et~al.}(1972)\citenamefont
  {Yamaguchi}, \citenamefont {Watanabe}, \citenamefont {Yamauchi},\ and\
  \citenamefont {Tomiyoshi}}]{yamaguchi1972}%
  \BibitemOpen
  \bibfield  {author} {\bibinfo {author} {\bibfnamefont {Y.}~\bibnamefont
  {Yamaguchi}}, \bibinfo {author} {\bibfnamefont {H.}~\bibnamefont {Watanabe}},
  \bibinfo {author} {\bibfnamefont {H.}~\bibnamefont {Yamauchi}},\ and\
  \bibinfo {author} {\bibfnamefont {S.}~\bibnamefont {Tomiyoshi}},\ }\bibfield
  {title} {\bibinfo {title} {Neutron {{Diffraction Study}} of
  {{Cr}}{\textsubscript{2}} {{As}}},\ }\href
  {https://doi.org/10.1143/JPSJ.32.958} {\bibfield  {journal} {\bibinfo
  {journal} {Journal of the Physical Society of Japan}\ }\textbf {\bibinfo
  {volume} {32}},\ \bibinfo {pages} {958} (\bibinfo {year} {1972})}\BibitemShut
  {NoStop}%
\bibitem [{\citenamefont {Katsuraki}\ and\ \citenamefont
  {Achiwa}(1966)}]{katsuraki1966}%
  \BibitemOpen
  \bibfield  {author} {\bibinfo {author} {\bibfnamefont {H.}~\bibnamefont
  {Katsuraki}}\ and\ \bibinfo {author} {\bibfnamefont {N.}~\bibnamefont
  {Achiwa}},\ }\bibfield  {title} {\bibinfo {title} {The {{Magnetic Structure}}
  of {{Fe}}{\textsubscript{2}} {{As}}},\ }\href
  {https://doi.org/10.1143/JPSJ.21.2238} {\bibfield  {journal} {\bibinfo
  {journal} {Journal of the Physical Society of Japan}\ }\textbf {\bibinfo
  {volume} {21}},\ \bibinfo {pages} {2238} (\bibinfo {year}
  {1966})}\BibitemShut {NoStop}%
\bibitem [{\citenamefont {Thakur}\ \emph {et~al.}(2014)\citenamefont {Thakur},
  \citenamefont {Haque}, \citenamefont {Gupta},\ and\ \citenamefont
  {Ganguli}}]{thakur2014}%
  \BibitemOpen
  \bibfield  {author} {\bibinfo {author} {\bibfnamefont {G.~S.}\ \bibnamefont
  {Thakur}}, \bibinfo {author} {\bibfnamefont {Z.}~\bibnamefont {Haque}},
  \bibinfo {author} {\bibfnamefont {L.~C.}\ \bibnamefont {Gupta}},\ and\
  \bibinfo {author} {\bibfnamefont {A.~K.}\ \bibnamefont {Ganguli}},\
  }\bibfield  {title} {\bibinfo {title} {{{CuFeAs}}: {{A New Member}} in the
  111-{{Family}} of {{Iron-Pnictides}}},\ }\href
  {https://doi.org/10.7566/JPSJ.83.054706} {\bibfield  {journal} {\bibinfo
  {journal} {Journal of the Physical Society of Japan}\ }\textbf {\bibinfo
  {volume} {83}},\ \bibinfo {pages} {054706} (\bibinfo {year}
  {2014})}\BibitemShut {NoStop}%
\bibitem [{\citenamefont {Wilkinson}\ \emph {et~al.}(1957)\citenamefont
  {Wilkinson}, \citenamefont {Gingrich},\ and\ \citenamefont
  {Shull}}]{wilkinson1957}%
  \BibitemOpen
  \bibfield  {author} {\bibinfo {author} {\bibfnamefont {M.}~\bibnamefont
  {Wilkinson}}, \bibinfo {author} {\bibfnamefont {N.}~\bibnamefont
  {Gingrich}},\ and\ \bibinfo {author} {\bibfnamefont {C.}~\bibnamefont
  {Shull}},\ }\bibfield  {title} {\bibinfo {title} {The magnetic structure of
  {{Mn2Sb}}},\ }\href {https://doi.org/10.1016/0022-3697(57)90074-4} {\bibfield
   {journal} {\bibinfo  {journal} {Journal of Physics and Chemistry of Solids}\
  }\textbf {\bibinfo {volume} {2}},\ \bibinfo {pages} {289} (\bibinfo {year}
  {1957})}\BibitemShut {NoStop}%
\bibitem [{\citenamefont {Qian}\ \emph {et~al.}(2012)\citenamefont {Qian},
  \citenamefont {Lee}, \citenamefont {Hu}, \citenamefont {Wang}, \citenamefont
  {Kumar}, \citenamefont {Fang}, \citenamefont {Liu}, \citenamefont {Fobes},
  \citenamefont {Pham}, \citenamefont {Spinu}, \citenamefont {Wu},
  \citenamefont {Green}, \citenamefont {Lee},\ and\ \citenamefont
  {Mao}}]{qian2012}%
  \BibitemOpen
  \bibfield  {author} {\bibinfo {author} {\bibfnamefont {B.}~\bibnamefont
  {Qian}}, \bibinfo {author} {\bibfnamefont {J.}~\bibnamefont {Lee}}, \bibinfo
  {author} {\bibfnamefont {J.}~\bibnamefont {Hu}}, \bibinfo {author}
  {\bibfnamefont {G.~C.}\ \bibnamefont {Wang}}, \bibinfo {author}
  {\bibfnamefont {P.}~\bibnamefont {Kumar}}, \bibinfo {author} {\bibfnamefont
  {M.~H.}\ \bibnamefont {Fang}}, \bibinfo {author} {\bibfnamefont {T.~J.}\
  \bibnamefont {Liu}}, \bibinfo {author} {\bibfnamefont {D.}~\bibnamefont
  {Fobes}}, \bibinfo {author} {\bibfnamefont {H.}~\bibnamefont {Pham}},
  \bibinfo {author} {\bibfnamefont {L.}~\bibnamefont {Spinu}}, \bibinfo
  {author} {\bibfnamefont {X.~S.}\ \bibnamefont {Wu}}, \bibinfo {author}
  {\bibfnamefont {M.}~\bibnamefont {Green}}, \bibinfo {author} {\bibfnamefont
  {S.~H.}\ \bibnamefont {Lee}},\ and\ \bibinfo {author} {\bibfnamefont {Z.~Q.}\
  \bibnamefont {Mao}},\ }\bibfield  {title} {\bibinfo {title} {Ferromagnetism
  in {{CuFeSb}}: {{Evidence}} of competing magnetic interactions in iron-based
  superconductors},\ }\href {https://doi.org/10.1103/PhysRevB.85.144427}
  {\bibfield  {journal} {\bibinfo  {journal} {Physical Review B}\ }\textbf
  {\bibinfo {volume} {85}},\ \bibinfo {pages} {144427} (\bibinfo {year}
  {2012})}\BibitemShut {NoStop}%
\bibitem [{\citenamefont {Yamaguchi}\ \emph {et~al.}(1999)\citenamefont
  {Yamaguchi}, \citenamefont {Ohoyama}, \citenamefont {Kanouchi}, \citenamefont
  {Ohashi},\ and\ \citenamefont {Ishimoto}}]{yamaguchi1999}%
  \BibitemOpen
  \bibfield  {author} {\bibinfo {author} {\bibfnamefont {Y.}~\bibnamefont
  {Yamaguchi}}, \bibinfo {author} {\bibfnamefont {K.}~\bibnamefont {Ohoyama}},
  \bibinfo {author} {\bibfnamefont {T.}~\bibnamefont {Kanouchi}}, \bibinfo
  {author} {\bibfnamefont {M.}~\bibnamefont {Ohashi}},\ and\ \bibinfo {author}
  {\bibfnamefont {K.}~\bibnamefont {Ishimoto}},\ }\bibfield  {title} {\bibinfo
  {title} {Multi-magnetic-mode in ({{Cr1}}-{{xMnx}}){{2As}}},\ }\href
  {https://doi.org/10.1016/S0022-3697(99)00091-8} {\bibfield  {journal}
  {\bibinfo  {journal} {Journal of Physics and Chemistry of Solids}\ }\textbf
  {\bibinfo {volume} {60}},\ \bibinfo {pages} {1229} (\bibinfo {year}
  {1999})}\BibitemShut {NoStop}%
\bibitem [{\citenamefont {Shirai}\ \emph {et~al.}(1993)\citenamefont {Shirai},
  \citenamefont {Kawamoto},\ and\ \citenamefont {Motizuki}}]{shirai1993}%
  \BibitemOpen
  \bibfield  {author} {\bibinfo {author} {\bibfnamefont {M.}~\bibnamefont
  {Shirai}}, \bibinfo {author} {\bibfnamefont {T.}~\bibnamefont {Kawamoto}},\
  and\ \bibinfo {author} {\bibfnamefont {K.}~\bibnamefont {Motizuki}},\
  }\bibfield  {title} {\bibinfo {title} {{{THE ELECTRONIC BAND STRUCTURES FOR
  AN ANTIFERROMAGNETIC STATE OF Cu}}{\textsubscript{2}} {{Sb-TYPE INTERMETALLIC
  COMPOUND Cr}}{\textsubscript{2}} {{As}}},\ }\href
  {https://doi.org/10.1142/S0217979293001621} {\bibfield  {journal} {\bibinfo
  {journal} {International Journal of Modern Physics B}\ }\textbf {\bibinfo
  {volume} {07}},\ \bibinfo {pages} {770} (\bibinfo {year} {1993})}\BibitemShut
  {NoStop}%
\bibitem [{\citenamefont {Zhang}\ \emph {et~al.}(2013)\citenamefont {Zhang},
  \citenamefont {Brgoch},\ and\ \citenamefont {Miller}}]{Zhang2013}%
  \BibitemOpen
  \bibfield  {author} {\bibinfo {author} {\bibfnamefont {Y.}~\bibnamefont
  {Zhang}}, \bibinfo {author} {\bibfnamefont {J.}~\bibnamefont {Brgoch}},\ and\
  \bibinfo {author} {\bibfnamefont {G.~J.}\ \bibnamefont {Miller}},\ }\bibfield
   {title} {\bibinfo {title} {Magnetic {{Ordering}} in {{Tetragonal}} 3d
  {{Metal Arsenides M}}{\textsubscript{2}} {{As}} ({{M}} = {{Cr}}, {{Mn}},
  {{Fe}}): {{An Ab Initio Investigation}}},\ }\href
  {https://doi.org/10.1021/ic3024716} {\bibfield  {journal} {\bibinfo
  {journal} {Inorganic Chemistry}\ }\textbf {\bibinfo {volume} {52}},\ \bibinfo
  {pages} {3013} (\bibinfo {year} {2013})}\BibitemShut {NoStop}%
\bibitem [{\citenamefont {Suzuki}\ \emph {et~al.}(1992)\citenamefont {Suzuki},
  \citenamefont {Shirai},\ and\ \citenamefont {Motizuki}}]{suzuki1992}%
  \BibitemOpen
  \bibfield  {author} {\bibinfo {author} {\bibfnamefont {M.}~\bibnamefont
  {Suzuki}}, \bibinfo {author} {\bibfnamefont {M.}~\bibnamefont {Shirai}},\
  and\ \bibinfo {author} {\bibfnamefont {K.}~\bibnamefont {Motizuki}},\
  }\bibfield  {title} {\bibinfo {title} {Electronic band structure in the
  ferrimagnetic state of {{Mn}}{\textsubscript{2}} {{Sb}}},\ }\href
  {https://doi.org/10.1088/0953-8984/4/1/008} {\bibfield  {journal} {\bibinfo
  {journal} {Journal of Physics: Condensed Matter}\ }\textbf {\bibinfo {volume}
  {4}},\ \bibinfo {pages} {L33} (\bibinfo {year} {1992})}\BibitemShut {NoStop}%
\bibitem [{\citenamefont {Parizek}\ and\ \citenamefont
  {Zelezny}(2026)}]{aiida_openmx_plugin_2026}%
  \BibitemOpen
  \bibfield  {author} {\bibinfo {author} {\bibfnamefont {V.}~\bibnamefont
  {Parizek}}\ and\ \bibinfo {author} {\bibfnamefont {J.}~\bibnamefont
  {Zelezny}},\ }\href@noop {} {\bibinfo {title} {Aiida openmx plugin}},\
  \bibinfo {howpublished}
  {\url{https://github.com/zeleznyj-htp/aiida-plugin-openmx}} (\bibinfo {year}
  {2026})\BibitemShut {NoStop}%
\bibitem [{\citenamefont {Fruchart}(2005)}]{Fruchart2005}%
  \BibitemOpen
  \bibfield  {author} {\bibinfo {author} {\bibfnamefont {D.}~\bibnamefont
  {Fruchart}},\ }\bibfield  {title} {\bibinfo {title} {Magnetic couplings in
  mm'x compounds of cu2sb type of structure (m,m'=d-metals, x=p-elements as p,
  as, sb, …)},\ }\href
  {https://doi.org/https://doi.org/10.1016/j.solidstatesciences.2004.11.023}
  {\bibfield  {journal} {\bibinfo  {journal} {Solid State Sciences}\ }\textbf
  {\bibinfo {volume} {7}},\ \bibinfo {pages} {767} (\bibinfo {year} {2005})},\
  \bibinfo {note} {a tribute to Erwin Felix Bertaut}\BibitemShut {NoStop}%
\bibitem [{\citenamefont {Pearson}(1985)}]{Pearson1985Cu2Sb}%
  \BibitemOpen
  \bibfield  {author} {\bibinfo {author} {\bibfnamefont {W.~B.}\ \bibnamefont
  {Pearson}},\ }\bibfield  {title} {\bibinfo {title} {The {Cu2Sb} and related
  structures},\ }\href {https://doi.org/10.1524/zkri.1985.171.1-2.23}
  {\bibfield  {journal} {\bibinfo  {journal} {Zeitschrift f{\"u}r
  Kristallographie}\ }\textbf {\bibinfo {volume} {171}},\ \bibinfo {pages} {23}
  (\bibinfo {year} {1985})}\BibitemShut {NoStop}%
\bibitem [{\citenamefont {Nuss}\ \emph {et~al.}(2006)\citenamefont {Nuss},
  \citenamefont {Wedig},\ and\ \citenamefont {Jansen}}]{Nuss2006Fe2As}%
  \BibitemOpen
  \bibfield  {author} {\bibinfo {author} {\bibfnamefont {J.}~\bibnamefont
  {Nuss}}, \bibinfo {author} {\bibfnamefont {U.}~\bibnamefont {Wedig}},\ and\
  \bibinfo {author} {\bibfnamefont {M.}~\bibnamefont {Jansen}},\ }\bibfield
  {title} {\bibinfo {title} {Geometric variations and electron localizations in
  intermetallics: Pbfcl type compounds},\ }\href
  {https://doi.org/10.1524/zkri.2006.221.5-7.554} {\bibfield  {journal}
  {\bibinfo  {journal} {Zeitschrift f{\"u}r Kristallographie - Crystalline
  Materials}\ }\textbf {\bibinfo {volume} {221}},\ \bibinfo {pages} {554}
  (\bibinfo {year} {2006})}\BibitemShut {NoStop}%
\bibitem [{\citenamefont {Wadley}\ \emph {et~al.}(2015)\citenamefont {Wadley},
  \citenamefont {Hills}, \citenamefont {Shahedkhah}, \citenamefont {Edmonds},
  \citenamefont {Campion}, \citenamefont {Nov{\'a}k}, \citenamefont
  {Ouladdiaf}, \citenamefont {Khalyavin}, \citenamefont {Langridge},
  \citenamefont {Saidl}, \citenamefont {Nemec}, \citenamefont {Rushforth},
  \citenamefont {Gallagher}, \citenamefont {Dhesi}, \citenamefont
  {Maccherozzi}, \citenamefont {{\v Z}elezn{\'y}},\ and\ \citenamefont
  {Jungwirth}}]{Wadley2015}%
  \BibitemOpen
  \bibfield  {author} {\bibinfo {author} {\bibfnamefont {P.}~\bibnamefont
  {Wadley}}, \bibinfo {author} {\bibfnamefont {V.}~\bibnamefont {Hills}},
  \bibinfo {author} {\bibfnamefont {M.~R.}\ \bibnamefont {Shahedkhah}},
  \bibinfo {author} {\bibfnamefont {K.~W.}\ \bibnamefont {Edmonds}}, \bibinfo
  {author} {\bibfnamefont {R.~P.}\ \bibnamefont {Campion}}, \bibinfo {author}
  {\bibfnamefont {V.}~\bibnamefont {Nov{\'a}k}}, \bibinfo {author}
  {\bibfnamefont {B.}~\bibnamefont {Ouladdiaf}}, \bibinfo {author}
  {\bibfnamefont {D.}~\bibnamefont {Khalyavin}}, \bibinfo {author}
  {\bibfnamefont {S.}~\bibnamefont {Langridge}}, \bibinfo {author}
  {\bibfnamefont {V.}~\bibnamefont {Saidl}}, \bibinfo {author} {\bibfnamefont
  {P.}~\bibnamefont {Nemec}}, \bibinfo {author} {\bibfnamefont {A.~W.}\
  \bibnamefont {Rushforth}}, \bibinfo {author} {\bibfnamefont {B.~L.}\
  \bibnamefont {Gallagher}}, \bibinfo {author} {\bibfnamefont {S.~S.}\
  \bibnamefont {Dhesi}}, \bibinfo {author} {\bibfnamefont {F.}~\bibnamefont
  {Maccherozzi}}, \bibinfo {author} {\bibfnamefont {J.}~\bibnamefont {{\v
  Z}elezn{\'y}}},\ and\ \bibinfo {author} {\bibfnamefont {T.}~\bibnamefont
  {Jungwirth}},\ }\bibfield  {title} {\bibinfo {title} {{Antiferromagnetic
  structure in tetragonal CuMnAs thin films}},\ }\href@noop {} {\bibfield
  {journal} {\bibinfo  {journal} {Scientific Reports}\ }\textbf {\bibinfo
  {volume} {5}},\ \bibinfo {eid} {17079} (\bibinfo {year} {2015})}\BibitemShut
  {NoStop}%
\bibitem [{\citenamefont {Tobola}\ \emph {et~al.}(2001)\citenamefont {Tobola},
  \citenamefont {Bacmann}, \citenamefont {Fruchart}, \citenamefont {Wolfers},
  \citenamefont {Kaprzyk},\ and\ \citenamefont {Koumina}}]{Tobola2001}%
  \BibitemOpen
  \bibfield  {author} {\bibinfo {author} {\bibfnamefont {J.}~\bibnamefont
  {Tobola}}, \bibinfo {author} {\bibfnamefont {M.}~\bibnamefont {Bacmann}},
  \bibinfo {author} {\bibfnamefont {D.}~\bibnamefont {Fruchart}}, \bibinfo
  {author} {\bibfnamefont {P.}~\bibnamefont {Wolfers}}, \bibinfo {author}
  {\bibfnamefont {S.}~\bibnamefont {Kaprzyk}},\ and\ \bibinfo {author}
  {\bibfnamefont {A.-A.}\ \bibnamefont {Koumina}},\ }\bibfield  {title}
  {\bibinfo {title} {Structure and magnetism in the polymorphous mnfeas},\
  }\href {https://doi.org/https://doi.org/10.1016/S0925-8388(00)01347-5}
  {\bibfield  {journal} {\bibinfo  {journal} {Journal of Alloys and Compounds}\
  }\textbf {\bibinfo {volume} {317-318}},\ \bibinfo {pages} {274} (\bibinfo
  {year} {2001})},\ \bibinfo {note} {the 13th International Conference on Solid
  Compounds of Transition Elements}\BibitemShut {NoStop}%
\bibitem [{\citenamefont {Naud}\ and\ \citenamefont {Priest}(1972)}]{Naud1972}%
  \BibitemOpen
  \bibfield  {author} {\bibinfo {author} {\bibfnamefont {J.}~\bibnamefont
  {Naud}}\ and\ \bibinfo {author} {\bibfnamefont {P.}~\bibnamefont {Priest}},\
  }\bibfield  {title} {\bibinfo {title} {Contribution a l'etude du systeme
  cuivre-arsenic},\ }\href
  {https://doi.org/https://doi.org/10.1016/0025-5408(72)90128-6} {\bibfield
  {journal} {\bibinfo  {journal} {Materials Research Bulletin}\ }\textbf
  {\bibinfo {volume} {7}},\ \bibinfo {pages} {783} (\bibinfo {year}
  {1972})}\BibitemShut {NoStop}%
\bibitem [{\citenamefont {Nuss}\ and\ \citenamefont {Jansen}(2002)}]{Nuss2002}%
  \BibitemOpen
  \bibfield  {author} {\bibinfo {author} {\bibfnamefont {J.}~\bibnamefont
  {Nuss}}\ and\ \bibinfo {author} {\bibfnamefont {M.}~\bibnamefont {Jansen}},\
  }\bibfield  {title} {\bibinfo {title} {Zur abgrenzung der {PbFCl}- und
  {Cu2Sb}-strukturfamilien: Neubestimmung und verfeinerung der
  kristallstrukturen von {CuMgSb}, {Cu2Sb} und {CuMgAs}},\ }\href
  {https://doi.org/10.1002/1521-3749(200206)628:5<1152::AID-ZAAC1152>3.0.CO;2-1}
  {\bibfield  {journal} {\bibinfo  {journal} {Zeitschrift f{\"u}r anorganische
  und allgemeine Chemie}\ }\textbf {\bibinfo {volume} {628}},\ \bibinfo {pages}
  {1152} (\bibinfo {year} {2002})}\BibitemShut {NoStop}%
\bibitem [{\citenamefont {Kamusella}\ \emph {et~al.}(2017)\citenamefont
  {Kamusella}, \citenamefont {Klauss}, \citenamefont {Thakur}, \citenamefont
  {Haque}, \citenamefont {Gupta}, \citenamefont {Ganguli}, \citenamefont
  {Kraft}, \citenamefont {Burkhardt}, \citenamefont {Rosner}, \citenamefont
  {Luetkens}, \citenamefont {Lynn},\ and\ \citenamefont
  {Zhao}}]{Kamusella2017}%
  \BibitemOpen
  \bibfield  {author} {\bibinfo {author} {\bibfnamefont {S.}~\bibnamefont
  {Kamusella}}, \bibinfo {author} {\bibfnamefont {H.-H.}\ \bibnamefont
  {Klauss}}, \bibinfo {author} {\bibfnamefont {G.~S.}\ \bibnamefont {Thakur}},
  \bibinfo {author} {\bibfnamefont {Z.}~\bibnamefont {Haque}}, \bibinfo
  {author} {\bibfnamefont {L.~C.}\ \bibnamefont {Gupta}}, \bibinfo {author}
  {\bibfnamefont {A.~K.}\ \bibnamefont {Ganguli}}, \bibinfo {author}
  {\bibfnamefont {I.}~\bibnamefont {Kraft}}, \bibinfo {author} {\bibfnamefont
  {U.}~\bibnamefont {Burkhardt}}, \bibinfo {author} {\bibfnamefont
  {H.}~\bibnamefont {Rosner}}, \bibinfo {author} {\bibfnamefont
  {H.}~\bibnamefont {Luetkens}}, \bibinfo {author} {\bibfnamefont {J.~W.}\
  \bibnamefont {Lynn}},\ and\ \bibinfo {author} {\bibfnamefont
  {Y.}~\bibnamefont {Zhao}},\ }\bibfield  {title} {\bibinfo {title} {Magnetism
  and site exchange in {CuFeAs} and {CuFeSb}: A microscopic and theoretical
  investigation},\ }\href {https://doi.org/10.1103/PhysRevB.95.094415}
  {\bibfield  {journal} {\bibinfo  {journal} {Physical Review B}\ }\textbf
  {\bibinfo {volume} {95}},\ \bibinfo {pages} {094415} (\bibinfo {year}
  {2017})}\BibitemShut {NoStop}%
\bibitem [{\citenamefont {\v{Z}elezn\'y}(2026)}]{ZELEZNY2026}%
  \BibitemOpen
  \bibfield  {author} {\bibinfo {author} {\bibfnamefont {J.}~\bibnamefont
  {\v{Z}elezn\'y}},\ }\bibfield  {title} {\bibinfo {title} {Symmetr: a python
  package for determining symmetry properties of crystals},\ }\href
  {https://doi.org/https://doi.org/10.1016/j.cpc.2026.110309} {\bibfield
  {journal} {\bibinfo  {journal} {Computer Physics Communications}\ ,\ \bibinfo
  {pages} {110309}} (\bibinfo {year} {2026})}\BibitemShut {NoStop}%
\bibitem [{\citenamefont {\v{S}mejkal}\ \emph {et~al.}(2022)\citenamefont
  {\v{S}mejkal}, \citenamefont {Sinova},\ and\ \citenamefont
  {Jungwirth}}]{Smejkal2022}%
  \BibitemOpen
  \bibfield  {author} {\bibinfo {author} {\bibfnamefont {L.}~\bibnamefont
  {\v{S}mejkal}}, \bibinfo {author} {\bibfnamefont {J.}~\bibnamefont
  {Sinova}},\ and\ \bibinfo {author} {\bibfnamefont {T.}~\bibnamefont
  {Jungwirth}},\ }\bibfield  {title} {\bibinfo {title} {Emerging research
  landscape of altermagnetism},\ }\href
  {https://doi.org/10.1103/PhysRevX.12.040501} {\bibfield  {journal} {\bibinfo
  {journal} {Phys. Rev. X}\ }\textbf {\bibinfo {volume} {12}},\ \bibinfo
  {pages} {040501} (\bibinfo {year} {2022})}\BibitemShut {NoStop}%
\bibitem [{\citenamefont {Ozaki}(2003)}]{Ozaki2003}%
  \BibitemOpen
  \bibfield  {author} {\bibinfo {author} {\bibfnamefont {T.}~\bibnamefont
  {Ozaki}},\ }\bibfield  {title} {\bibinfo {title} {Variationally optimized
  atomic orbitals for large-scale electronic structures},\ }\href
  {https://doi.org/10.1103/PhysRevB.67.155108} {\bibfield  {journal} {\bibinfo
  {journal} {Phys. Rev. B}\ }\textbf {\bibinfo {volume} {67}},\ \bibinfo
  {pages} {155108} (\bibinfo {year} {2003})}\BibitemShut {NoStop}%
\end{thebibliography}%

\end{document}